\documentclass[useAMS,usenatbib]{mn2e}
\usepackage{epsfig}
\def\mnras{{MNRAS}}
\def\aj{{AJ}}

\def\apj{{ApJ}}
\def\apjs{{ApJS}}
\def\apjl{{ApJ Lett.}}
\def\aap{{A\&A}}
\def\araa{{ARA\&A}}

\def\nat{{Nature}}

\def\pasp{{P.A.S.P.}}
\def\aap{{AAP}}

\title[]
 {Globular Cluster Systems of Early-type Galaxies in Low-density Environments\thanks{Based on observations
made with the NASA/ESA {\it Hubble Space Telescope},
obtained at the Space Telescope Science Institute,
which is operated by the Association of Universities
for Research in Astronomy, Inc., under NASA contract
NAS5-26555.}}

\author[]
  {J.~Cho,$^{1,2}$
  R.M.~Sharples,$^2$
  J.P.~Blakeslee,$^3$ S.E.~Zepf,$^4$
  A.~Kundu,$^5$ 
  H.-S.~Kim$^{1,6}$ and 
  \newauthor S.-J.~Yoon$^{1,6}$
  \thanks{E-mail: sjyoon@galaxy.yonsei.ac.kr}\\ \\
  $^1$Department of Astronomy, Yonsei University, Seoul, 120-749, Korea\\
  $^2$Department of Physics, University of Durham, South Road, Durham, DH1 3LE\\
  $^3$Herzberg Institute of Astrophysics, National Research Council of Canada, Victoria, BC V9E 2E7, Canada\\
  $^4$Department of Physics and Astronomy, Michigan State University, East Lansing, MI 48824, USA\\
  $^5$Eureka Scientific, 2452 Delmer Street Suite 100, Oakland, CA 94602-3017, USA\\
  $^6$Center for Galaxy Evolution Research, Yonsei University, Seoul, 120-749, Korea
  }

\date{Accepted 2012 March 2. Received 2012 March 1; in original form 2011 November 15}

\pagerange{\pageref{1}--\pageref{20}} \pubyear{2012}

\def\LaTeX{L\kern-.36em\raise.3ex\hbox{a}\kern-.15em
    T\kern-.1667em\lower.7ex\hbox{E}\kern-.125emX}

\begin{document}

\label{firstpage}

\maketitle

\begin{abstract}
Deep images of 10 early-type galaxies in low-density environments have
been obtained with the Advanced Camera for Surveys (ACS) on the Hubble
Space Telescope. The global properties of the globular cluster (GC)
systems of the galaxies have been derived in order to
investigate the role of the environment in galaxy formation and
evolution. Using the ACS Virgo Cluster Survey (ACSVCS) as a high-density counterpart, the similarities and differences between the GC
properties in high- and low-density environments are presented. We
find a strong correlation of the GC mean colours and the degree of
colour bimodality with the host galaxy luminosity in low-density environments, in good
agreement with high-density environments. In contrast, the GC mean colours
at a given host luminosity are somewhat bluer ($\Delta(g-z)\sim0.05$) than
those for cluster galaxies, indicating more metal-poor ($\Delta[Fe/H]\sim0.10-0.15$) and/or younger ($\Delta age>2$ Gyr) GC systems than those in
dense environments. Furthermore, with decreasing host luminosity, the
colour bimodality disappears faster, when compared to galaxies in
cluster environments. Our results suggest that: (1) in both
high- {\it and} low-density environments, the mass of the host galaxy has the dominant
effect on GC system properties, (2)  the local environment has only a secondary effect on the history of GC system formation, (3) GC formation must be governed by common physical processes across a range of environments.

\end{abstract}

\begin{keywords}
galaxies: elliptical and lenticular, cD --- galaxies: evolution --- galaxies: formation
 
\end{keywords}

\section{Introduction}
The study of extragalactic globular cluster (GC) systems is now well-established as a powerful probe of the formation histories of elliptical galaxies (e.g. \citealt{wes04}; \citealt{bro06}). Because each cluster is essentially of a single age, a single metallicity, and a simple stellar population (SSP), the colours and line strengths of individual clusters are more easily interpreted using evolutionary synthesis stellar population models than is the integrated starlight from the spheroidal component. The distribution of the aforementioned properties over the entire cluster population also allows for a unique examination of the relative importance of cluster formation episodes as a function of age and metallicity. Furthermore, the well-documented correlations between the properties of GC systems and those of their host galaxies (e.g. \citealt{ash98}; \citealt{lar01}; \citealt{bro06}; \citealt{pen06a}, hereafter PJC06) indicate that a close connection must exist between the formation of the diffuse stellar component of a galaxy and that of its GC system.

\begin{table*}
%\begin{center}
 \begin{minipage}{130mm}
% \begin{minipage}{110mm}
  \caption{The galaxy properties}
  \begin{tabular}{ccccccccc}

\hline
 Name & $M_B$ & RA (J2000) & DEC (J2000) & Type & m-M & $B_T$ & $E(B-V)$ & Density  \\
  &  & (h:m:s) & (\degr:\arcmin:\arcsec) &  &  & $ $ &  & ($Mpc^{3}$)  \\
\hline
NGC 3818 & $-$20.33 & 11:41:57.36 & $-$06:09:20.4 &   E5 & 32.80$\pm$0.61 & 12.47 &  0.036 & 0.20 \\
NGC 7173 & $-$19.96 & 22:02:03.19 & $-$31:58:25.3 & E+ p & 32.48$\pm$0.19 & 12.52 &  0.026 & 0.35 \\
NGC 1439 & $-$19.95 & 03:44:49.95 & $-$21:55:14.0 &   E1 & 32.13$\pm$0.15 & 12.18 &  0.030 & 0.45 \\
NGC 1426 & $-$19.65 & 03:42:49.11 & $-$22:06:30.1 &   E4 & 31.91$\pm$0.18 & 12.26 &  0.017 & 0.66 \\
NGC 3377 & $-$19.18 & 10:47:42.40 & +13:59:08.3 &   E5 & 30.25$\pm$0.09 & 11.07 &  0.034 & 0.49 \\
NGC 4033 & $-$19.11 & 12:00:34.74 & $-$17:50:33.4 &   E6 & 31.64$\pm$0.23 & 12.53 &  0.047 & 0.38 \\
NGC 1172 & $-$19.10 & 03:01:36.05 & $-$14:50:11.7 &   E+ & 31.66$\pm$0.20 & 12.56 &  0.064 & 0.28 \\
NGC 3156 & $-$18.84 & 10:12:41.25 & +03:07:45.7 &   S0 & 31.75$\pm$0.14 & 12.91 &  0.034 & 0.20 \\
NGC 3073 & $-$18.78 & 10:00:52.08 & +55:37:07.8 & SAB0 & 32.64$\pm$0.93 & 13.86 &  0.010 & 0.28 \\
IC 2035 & $-$18.57 & 04:09:01.87 & $-$45:31:03.1 &   S0 & 31.18$\pm$0.15 & 12.61 &  0.013 & 0.16 \\
\hline

\end{tabular}  
\medskip  
%\vspace{10mm}

Col. 1: Galaxy name; Col. 2: Absolute B magnitude; Col. 3: RA; Col. 4: Dec; Col. 5: Morphological type; Col. 6: Distance modulus; Col. 7:
Total apparent B magnitude;  Col. 8: Galactic
reddening; Col. 9: Local density. 
The morphological type and apparent B magnitude are taken
from NED. The distance moduli are SBF values from \citet{ton01}  except for IC 2035
where the distance modulus was estimated using the redshift from NED. The Galactic
extinction is taken from \citet{sc98} and the local density was obtained from \citet{tul88}.

\end{minipage}
%\end{center}
\label{tab:gal_pro}
\end{table*}

Most of the extant studies on globular systems in early-type galaxies have concentrated on high-density environments where these morphological types are most prevalent. Such studies include the ACS Virgo Cluster Survey \citep{cot04}, the ACS Fornax Cluster Survey \citep{jor07a}, and the ACS Coma Cluster Survey  \citep{car08}. Semi-analytic models of galaxy formation (e.g. \citealt{bau98}; \citealt{bea02}) predict that the largest spread in galaxy properties such as metallicity and age should occur for early-type galaxies (L $<$ L$_\ast$) in low-density environments outside of rich clusters. Low-luminosity ellipticals also provide a critical test for current theories on the origin of bimodal colour distributions in GC systems. In some scenarios, clusters in the blue (low-metallicity) peak are primarily associated with the progenitor galaxy; red cluster populations form later via multi-phase collapse \citep{for97} or mergers \citep{ash92}. It has also been proposed by \citet{cot98} that red clusters represent the original population and blue clusters have been accreted from a lower metallicity dwarf galaxy population, in which case the low-luminosity ellipticals would be expected to contain any smaller population of red, metal-rich clusters \citep{kun01a}. In N-body simulations (e.g. \citealt{mur10}), metallicity bimodality arises from the history of galaxy assembly. For instance, early mergers are frequently involving relatively low mass protogalaxies, which preferentially produce low metallicity blue clusters. Late mergers are more infrequent but involve more massive galaxies and just a few late massive mergers can produce a significant number of red clusters. 
Alternatively, \citet{yoo06}, \citet{yoo11a}, and \citet{yoo11b} suggested that a broad, single-peaked metallicity distribution presumably as a result of continuous chemical enrichment can produce colour bimodality due to the non-linear metallicity-to-colour conversion, without invoking two distinct sub-populations. In fact, the study of GC populations in low-luminosity ellipticals may be the only way of distinguishing between the above models since the predictions from all of them are very similar for more massive galaxies (e.g. \citealt{kis01}). It is crucial to test these predictions for galaxies in both rich clusters and in the field where, for example, accretion processes may be by far less efficient.

\begin{figure*}
\begin{center}
\includegraphics[width=170mm]{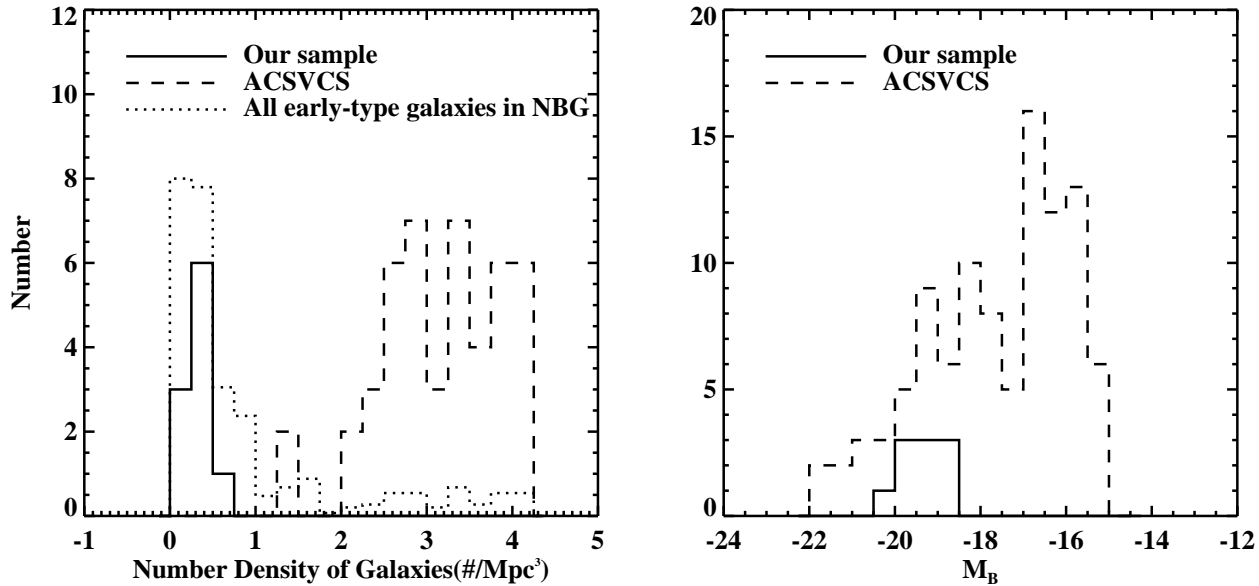}
\caption[]{Distribution of the local density and the absolute magnitude in B for our sample and the ACSVCS. {\it left panel}: The number densities are taken from the NBG \citep{tul88}. A clear separation in the environment between our sample (solid histogram) and the ACSVCS (dashed histogram) is shown. The distribution of all
early-type galaxies in the NBG \citep{tul88} is plotted with a dotted line on an arbitrary scale. {\it right panel}: The solid histogram represents the distribution of the absolute B magnitude for our sample (Table \ref{tab:gal_pro}), while the dashed histogram indicates that of the ACSVCS galaxies.}
  \label{fig:rho}
 \end{center} 
\end{figure*}

The current level of knowledge regarding GC populations in low-luminosity ellipticals is quite poor. \citet{kis98} concluded on the basis of limited ground-based data that the colours of their GC populations were consistent with a unimodal distribution, i.e., the colours were very different from those of luminous ellipticals. However, ground-based imaging is severely compromised by the need for large statistical background corrections. Compounding the problem is the proportionately smaller number of clusters associated with lower luminosity galaxies. Using compilations of archival WFPC2 HST data, \citet{geb99} and \citet{kun01a} both found evidence for possible bimodality in a small sample of lower luminosity ellipticals. However, the limited resolution and depth of the WFPC2 data produced discrepancies in the classification of individual galaxies. The properties of low-luminosity galaxies in high-density environments has been addressed by the ACS Virgo Cluster Survey (\citealt{pen06a}; \citealt{pen08}). However, such a sample is, by its nature, biased towards environments that are atypical of a general field population (Figure \ref{fig:rho}). A complementary field galaxy comparison sample selected from low-density environments is required to investigate the role of environmental effects on the physical processes that control the formation of GC systems and their host galaxies.

\begin{figure}
\begin{center}
\includegraphics[width=80mm]{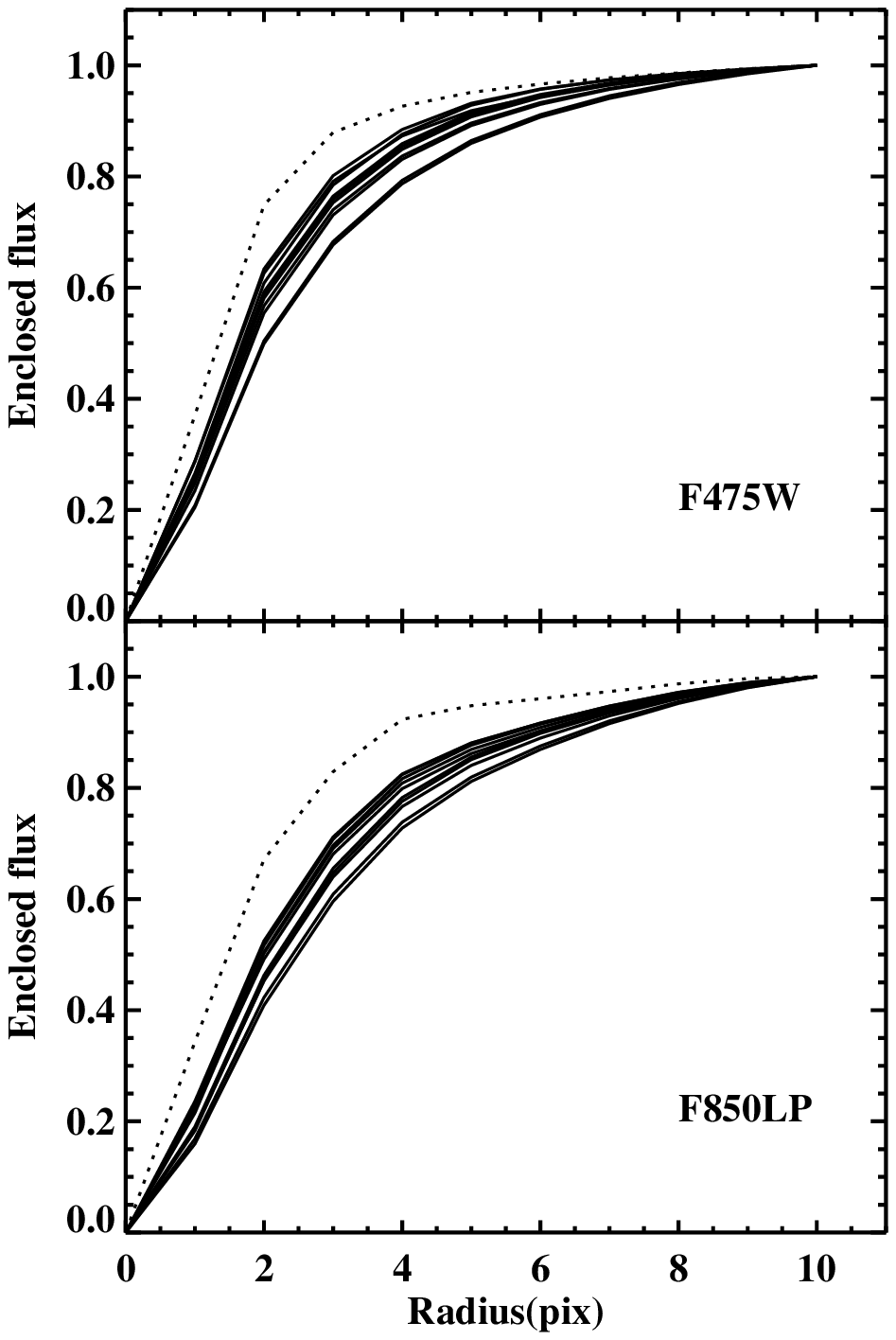}
      \caption[Growth curves of bright GCs in different sample galaxies]{Growth curves of bright GCs in the different sample galaxies. The dotted lines are growth curves for
       the ACS PSF. The GCs are clearly more extended than a point source. The most
       extended profile is from the closest galaxy (NGC 3377).}
     \label{fig:gc_gc}
\end{center}
\end{figure}

\section{Sample Selection and Observation}
In order to obtain a well-defined sample of early-type galaxies with low-luminosities in low-density environments, a complete sample of galaxies was first compiled from the Nearby Galaxy Catalogue (NBG) \citep{tul88} with the following parameters: $-5$ $\leq$ $H_{code}$ $\leq$ $-2$\footnote{This is a morphological type code. Elliptical and lenticular galaxies are included in this range.}, $-18.0$ $ <$ $M_B$ $ <$ $-19.5$\footnote{The absolute magnitudes of galaxies listed in the NBG scatter by $\sim$ 0.5 mag
compared to the values from the NASA Extragalactic Database (NED). We adopt the values from the NED for the presents analysis.}, distance $D \la$ 30 Mpc, Galactic latitude $|b| \ga 45 \deg$, and local density $\rho_0$ $<$ 1.0 $Mpc^{-3}$. 
\citet{tul88} determines the local density, $\rho_0$, on a three dimensional grid with a scale size of 0.5 Mpc after Gaussian smoothing, to estimate the contribution of each member of the galaxy population brighter than $M_B$ $<$ $-16$ mag. 
After a visual inspection of the Digitized Sky Survey images, 10 of the brightest (in apparent magnitude) targets with no nearby bright stars within the ACS field of view were selected. The properties of the final sample of early-type galaxies in low-density environments are listed in  Table \ref{tab:gal_pro} in descending order of absolute luminosity. The local densities of 47 galaxies from the ACSVCS sample are also available in the NBG, including all ACSVCS galaxies within the luminosity range of our sample. As shown in the left panel of Figure \ref{fig:rho}, there is not only a clear separation in the local density between our field galaxy sample and the ACSVCS sample, but the sample in low-density environments is also more representative of early-type galaxies in general. The right panel of Figure  \ref{fig:rho} shows the distribution in the absolute magnitude of our sample and the ACSVCS sample.

The observations (Program ID: 10554) were carried out with the
ACS Wide Field Camera (WFC) on the Hubble Space
Telescope during Cycle 14 (Oct 05 -- Sep 06). 
The WFC is composed of two 4K $\times$ 2K chips with a pixel scale of $0.05\arcsec/pix$ covering a field of view of 202$\arcsec$ $\times$ 202$\arcsec$. For each galaxy, 2 orbits were used to obtain images in two bands, F475W and F850LP, which correspond to Sloan filters {\it g\/} and {\it z\/}, respectively. These two filters were also used by the ACSVCS, which has the advantage of being able to compare results directly without the use of photometric transformations. In this work, a four-point line dithering pattern in which the telescope pointing shifts 5 $\times$ 60 pixels between sub-exposures was employed. Such a pattern allows the gap between the two chips to be filled and eliminates hot pixels during data processing. Because dithering also effectively removes cosmic rays, {\tt CR-SPLIT} was set to NO. The total exposure times were maximized by arranging 8 sub-exposures in the following way. In the first orbit, four sub-exposures of F475W and one sub-exposure of F850LP were allocated. In the second orbit, three sub-exposures of F850LP were obtained. We aim to achieve a S/N $\sim$ 30 at 50\% completeness, corresponding to estimated exposure times of 23 minutes in F475W and 53 minutes in F850LP. The actual total exposure times depend on the visibility of the target and were typically 1300 s and 3000 s for F475W and F850LP, respectively. The observation logs are listed in Table \ref{tab:obs_log}. For NGC 3156, the pointing and orientation were slightly adjusted so as to avoid adjacent bright stars.

\begin{table}
\begin{center}
% \begin{minipage}{126mm}
% \begin{minipage}{110mm}
  \caption{The observation logs\label{tab:obs_log}}
  \begin{tabular}{ccccc}
\hline
 Target & Date & Filter & Exp. Time & Dataset  \\
 & & & (sec) & \\
\hline
NGC 3818 & 2006-01-01 &  F475W & 1380 & J9CZ04010 \\
  &   & F850LP & 2987 & J9CZ04020 \\
NGC 7173 & 2006-05-16 &  F475W & 1375 & J9CZ10010 \\
  &   & F850LP & 3075 & J9CZ10020 \\
NGC 1439 & 2006-08-21 &  F475W & 1375 & J9CZ06010 \\
  &   & F850LP & 3023 & J9CZ06020 \\
NGC 1426 & 2006-08-21 &  F475W & 1375 & J9CZ09010 \\
  &   & F850LP & 3023 & J9CZ09020 \\
NGC 3377 & 2006-01-13 &  F475W & 1380 & J9CZ08010 \\
  &   & F850LP & 3005 & J9CZ08020 \\
NGC 4033 & 2006-01-04 &  F475W & 1380 & J9CZ07010 \\
  &   & F850LP & 3017 & J9CZ07020 \\
NGC 1172 & 2006-08-17 &  F475W & 1380 & J9CZ03010 \\
  &   & F850LP & 3005 & J9CZ03020 \\
NGC 3156 & 2005-10-30 &  F475W & 1380 & J9CZ02010 \\
  &   & F850LP & 2972 & J9CZ02020 \\
NGC 3073 & 2006-05-14 &  F475W & 1428 & J9CZ05010 \\
  &   & F850LP & 3490 & J9CZ05020 \\
 IC 2035 & 2006-04-28 &  F475W & 1380 & J9CZ01010 \\
  &   & F850LP & 3299 & J9CZ01020 \\
\hline
\end{tabular}
%\medskip
 
%\end{minipage}
\end{center}

\end{table}

\section{Data Reduction}

\subsection{Basic image reduction}

Each target has four sub-exposure images with different dithering positions in each filter. These raw images were reduced using the default ACS pipeline, {\it CALACS}, to produce bias subtracted and flat-field corrected images. The default processing was found to slightly overestimate the sky background because of the large contribution of galaxy light in the images. This resulted in negative background levels on the outskirts of a galaxy in the final drizzled images. An improved sky background was determined by adopting the minimum median value from four 200 pix $\times$ 200 pix corners of each sub-exposure. The four sub-exposure images were then combined and geometrically corrected with the {\it MultiDrizzle} package. The final drizzled image consists of a 4096 $\times$ 4096 pixel science image in units of electrons/s and an error map in the second extension that contains all error sources such as readout noise, dark current, and photon background noise.

\subsection{Galaxy light subtraction}
\label{galsub}

The smooth galaxy light of each target was fitted with elliptical isophotes using the {\it IRAF/ELLIPSE} routine. Fitted ellipse parameter tables were then employed in order to build a galaxy model image using the {\it IRAF/BMODEL} task.  The model galaxy was subtracted from the original image. It was found that subtracting the galaxy continuum light improves the efficiency of GC detection in the central regions of a galaxy. It is because that ellipse fitting does better job to model galaxy light in the central regions than {\it SExtractor}  \citep{ber96}, which models background (galaxy light in our case) using interpolation of mean/median values in a specified mesh size, thus the galaxy light in the centre may be underestimated.  

\begin{figure*}
\begin{center}
\includegraphics[width=153mm]{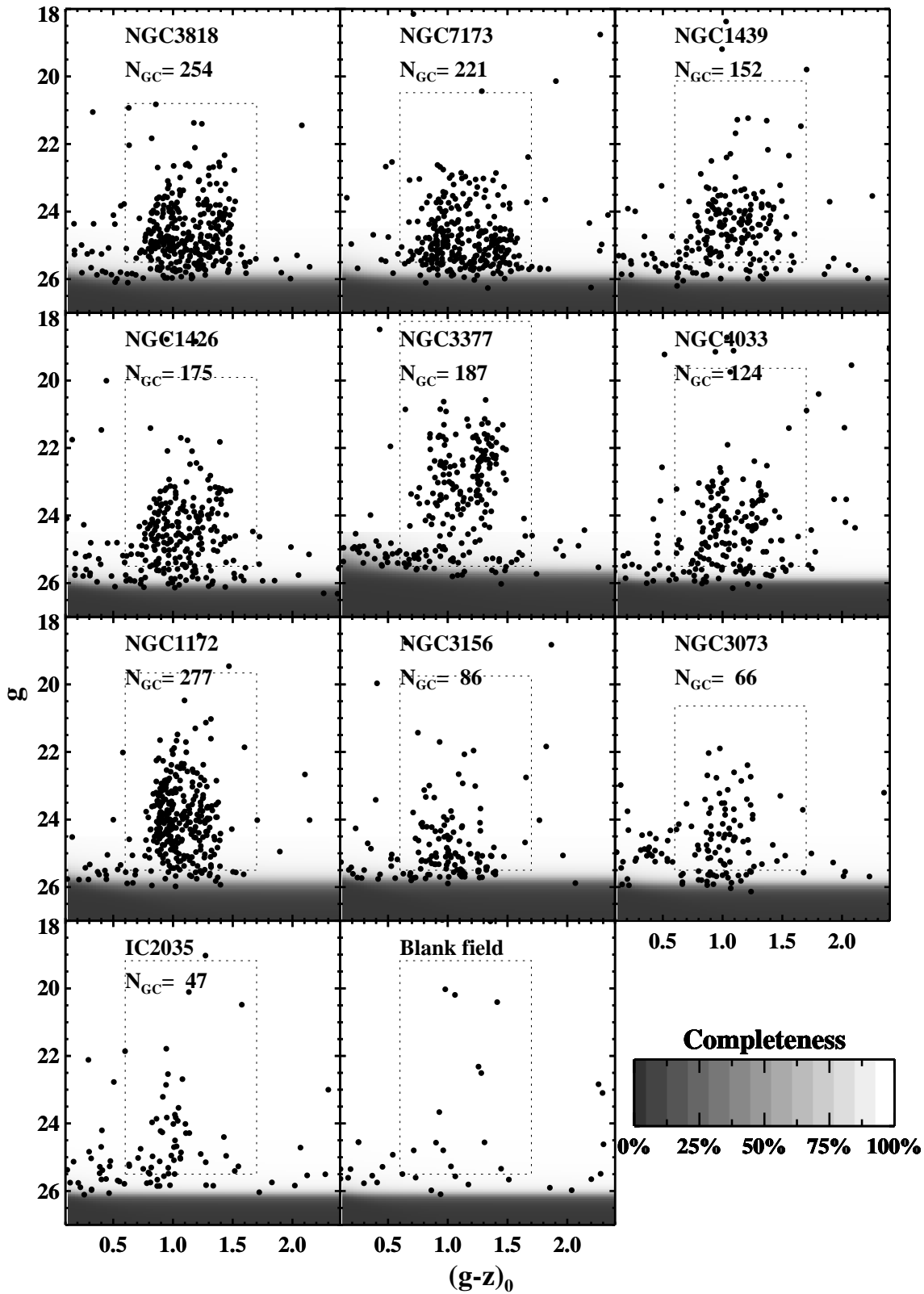}
       \caption[The colour-magnitude diagram of GC candidates]{The colour-magnitude diagram of GC candidates; the parent galaxies are ordered from most luminous (NGC 3818) to least luminous (IC 2035). The dotted box
       indicates the applied colour and the magnitude cut applied. The number of GCs in
       this box is denoted along with the galaxy name. The background grey scale
        represents the completeness at a given magnitude and colour. The
	bottom-center panel shows an example of contamination per field, after one
	out of six objects is randomly selected from 6 compiled blank fields.}
     \label{fig:cmd}
\end{center}
\end{figure*}

\subsection{Globular cluster detection and selection}
\label{gcdet}

{\it SExtractor} was run on the galaxy-light subtracted images so as to detect GC candidates with ${\tt DETECT\_THRESH} > 3\sigma$ above the background. In order to properly account for noise due to diffuse galaxy light, the error map from the drizzled images was used to create a weight image in {\it SExtractor}. Among the detected objects, those with ${\tt ELONGATION\/} > 2$ in either the F475W or F850LP band and very diffuse objects with ${\tt CLASS\_STAR\/} < 0.9$ in the F475W band were rejected. Objects were matched within a 2 pixel radial separation across the two bands. Some spurious detections that were found at the edge of the images were manually removed.

\subsection{Globular cluster photometry}
\label{sec:low_phot}

Once the list of GC candidates had been generated, aperture photometry was performed on the galaxy-subtracted images using the {\it IRAF/PHOT} package with 1 to 10 pixel radius apertures in one pixel steps. The background was estimated locally from a 10 to 20 pixel radius annulus. This locally estimated background has the advantage that the photometry of GCs is hardly affected by the global sky background estimation and the modeling of underlying galaxy light. Because GCs are marginally resolved by ACS/WFC at the distance of our sample galaxies, the aperture correction from stellar PSFs would not be applicable to our GCs. Instead, a representative GC model profile was constructed from several moderately bright GCs for each galaxy using the {\it IRAF/PSF} task. From this process, growth curves were extracted. As clearly shown in Figure \ref{fig:gc_gc}, the GCs are not point sources, which can be verified by the fact that the light profiles of the GCs in the closest galaxy, NGC 3377, are the most extended.

An aperture correction for the GCs in each galaxy was determined from the magnitude difference between the 3 pixel and 10 pixel radius apertures on each growth curve. This aperture correction was then applied to the 3 pixel radius aperture photometric measurements for all of the GC candidates. Such a process can be justified by the fact that the sizes of GCs are independent of their luminosity  \citep{jo05}. The final corrections from the 10 pixel radius to the total magnitudes are primarily a function of the ACS/WFC PSF and are 0.095 mag for F475W and 0.117 mag for F850LP; the zero-points of the ABmag scale are 26.068 and 24.862 for F475W and F850LP, respectively \citep{si05}. A correction for Galactic extinction was implemented based on the \citet{sc98} extinction map values (listed in Table~\ref{tab:gal_pro}) and the extinction ratios, $A_{F475W}=3.634E(B-V)$ and $A_{F850LP}=1.485E(B-V)$, from \citet{si05}. Hereafter, {\it g} and {\it z} magnitudes refer to the F475W and F850LP extinction-corrected total ABmag. Colour-magnitude diagrams for the GC candidates, in order of their host galaxy luminosity, are shown in Figure~\ref{fig:cmd}, with the most luminous galaxy in the top-left panel.

\subsection{Globular cluster completeness tests}

To test the effectiveness of the proposed GC detection and selection method, a list of 1000 artificial GCs with a uniform luminosity function and a spatial distribution in both filters was independently generated. These GCs were then added onto a galaxy subtracted image that contains all relevant noise sources such as the readout noise and poisson errors from sky and galaxy light. The artificial GCs were simulated using {\it IRAF/MKOBJECTS} with the representative GC light profile detailed in \S \ref{sec:low_phot}. The GC detection and selection procedures described in \S~\ref{gcdet} were then repeated and the initial input coordinates of the artificial GCs were matched with those of the recovered GCs. For a given magnitude bin, the number fractions of recovered GCs were calculated.

\begin{figure}
\begin{center}
\includegraphics[width=87mm]{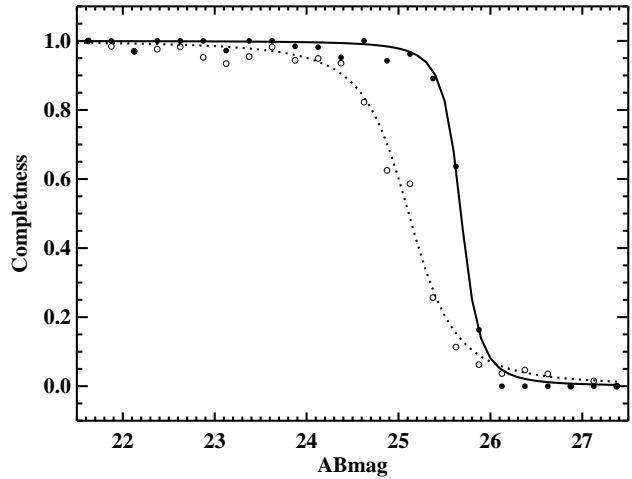}
\caption[The completeness of GC detections in NGC 3377]{The completeness of GC detections in NGC 3377. The filled and open circles denote the completeness of the $g$ and $z$ bands respectively. The solid and dotted lines are the fitted curves from Equation \ref{eq:compl}.}
\label{fig:compl}
\end{center}
\end{figure}

A completeness curve can often be expressed by the analytic function,
\begin{equation} \label{eq:compl}
f=\frac{1}{2}\Big(1-\frac{a(m-m_0)}{\sqrt{1+a^2(m-m_0)^2}}\Big) 
\end{equation} 
where  $m_0$ is a magnitude when  $f$ is 0.5 and $a$ controls how quickly $f$ declines (the larger the value of $a$, the steeper the transformation from 1 to 0) \citep{fle95}. An example of completeness in one of the galaxies is shown in Figure~\ref{fig:compl}. The fitted parameters of Equation~\ref{eq:compl} are listed in Table~\ref{tab:compl}. The completeness in each galaxy at a given magnitude and colour is displayed as a grey scale in Figure~\ref{fig:cmd}. It is calculated by multiplying the completeness in $g$ and $z$ over a grid of small boxes on the colour-magnitude plane. As seen in Figure~\ref{fig:compl} and Table~\ref{tab:compl}, the completeness levels in $g$ decline more steeply than in z. This is because the additional selection criterion in g, {\tt CLASS\_STAR}, rejects faint GCs more quickly due to the increasing uncertainty of {\tt CLASS\_STAR} in fainter GCs.

\begin{table}
\begin{center}
% \begin{minipage}{126mm}
% \begin{minipage}{130mm}
  \caption{The parameters of the completeness functions\label{tab:compl}}
  \begin{tabular}{ccccc}
\hline
Galaxy  & \multicolumn{2}{c}{F475W} & \multicolumn{2}{c}{F850LP} \\
   & $m_0$ & $a$ & $m_0$ & $a$ \\
\hline
NGC 3818 & 25.94 & 4.29 & 25.79 & 3.06 \\
NGC 7173 & 25.93 & 5.24 & 25.74 & 2.73 \\
NGC 1439 & 26.01 & 6.17 & 25.84 & 3.12 \\
NGC 1426 & 26.05 & 5.55 & 25.85 & 3.44 \\
NGC 3377 & 25.68 & 4.81 & 25.11 & 1.88 \\
NGC 4033 & 25.90 & 7.06 & 25.81 & 3.81 \\
NGC 1172 & 25.86 & 3.85 & 25.65 & 4.03 \\
NGC 3156 & 25.77 & 4.88 & 25.65 & 5.10 \\
NGC 3073 & 25.92 & 5.48 & 25.79 & 3.88 \\
IC 2035 & 26.10 & 6.77 & 26.02 & 4.85 \\
\hline
\end{tabular}
%\medskip
 
%\end{minipage}
\end{center}

\end{table}

 %\eqref{...} 
 
 \subsection{Contamination of background galaxies} 
 
 Although our selection criteria are quite strict, it is inevitable that some amount of contamination by background galaxies will be present (contamination by foreground stars is small at these magnitudes with our selection criteria). To estimate this contamination statistically, the ACS archive was searched for high Galactic latitude blank-fields (proposal ID: 9488) that were observed with the same filters and were at least as deep as our observations. In total 30 drizzled images from six blank fields were retrieved. Table \ref{tab:control} lists RA, DEC and the number of exposures of each band for the chosen blank fields. One field usually consists of 3-6 sub-exposures that were combined into one image by taking median values via {\it IRAF/IMCOMBINE}. The aforementioned GC detection and selection techniques were applied to these images. Since the observation conditions for the blank fields are not identical to ours (e.g. exposure time and dithering), this difference must be taken into account. It was found that some hot pixels were misclassified as GCs in the blank fields. These were removed by applying an additional classification, ${\tt FWHM\_IMAGE} >1.5$. In the bottom-center panel in Figure \ref{fig:cmd}, one out of every six objects in the 6 blank fields that passed the selection criteria is randomly chosen and plotted. Because the blank fields are deeper than ours, the completeness of each galaxy was applied so as to obtain the colour distribution and luminosity function of the contamination for each galaxy. These contamination profiles were used in the subsequent analysis. The typical contamination per field is $\sim$ 14 objects.

\begin{table}
\begin{center}
% \begin{minipage}{126mm}
% \begin{minipage}{130mm}
  \caption{Control fields \label{tab:control}}
  \begin{tabular}{ccc}
\hline
  RA (J2000) & DEC (J2000) & Number of exposures \\
 &	&	 for each filter\\
\hline
14h18m28.76s & +24d59m21.95s & 5 \\
10h34m05.03s &  +58d57m50.60s & 6\\
22h22m55.53s & $-$72d23m12.21s & 6 \\
12h43m30.12s & +11d49m20.62s & 6\\
12h10m53.80s & +39d14m06.24s & 4\\
11h13m33.33s & +22d15m41.22s & 3\\
\hline
\end{tabular}
%\medskip
 
%\end{minipage}
\end{center}

\end{table}

\section{Results}
\subsection{Colour distributions}

\begin{figure*}
\includegraphics[width=170mm]{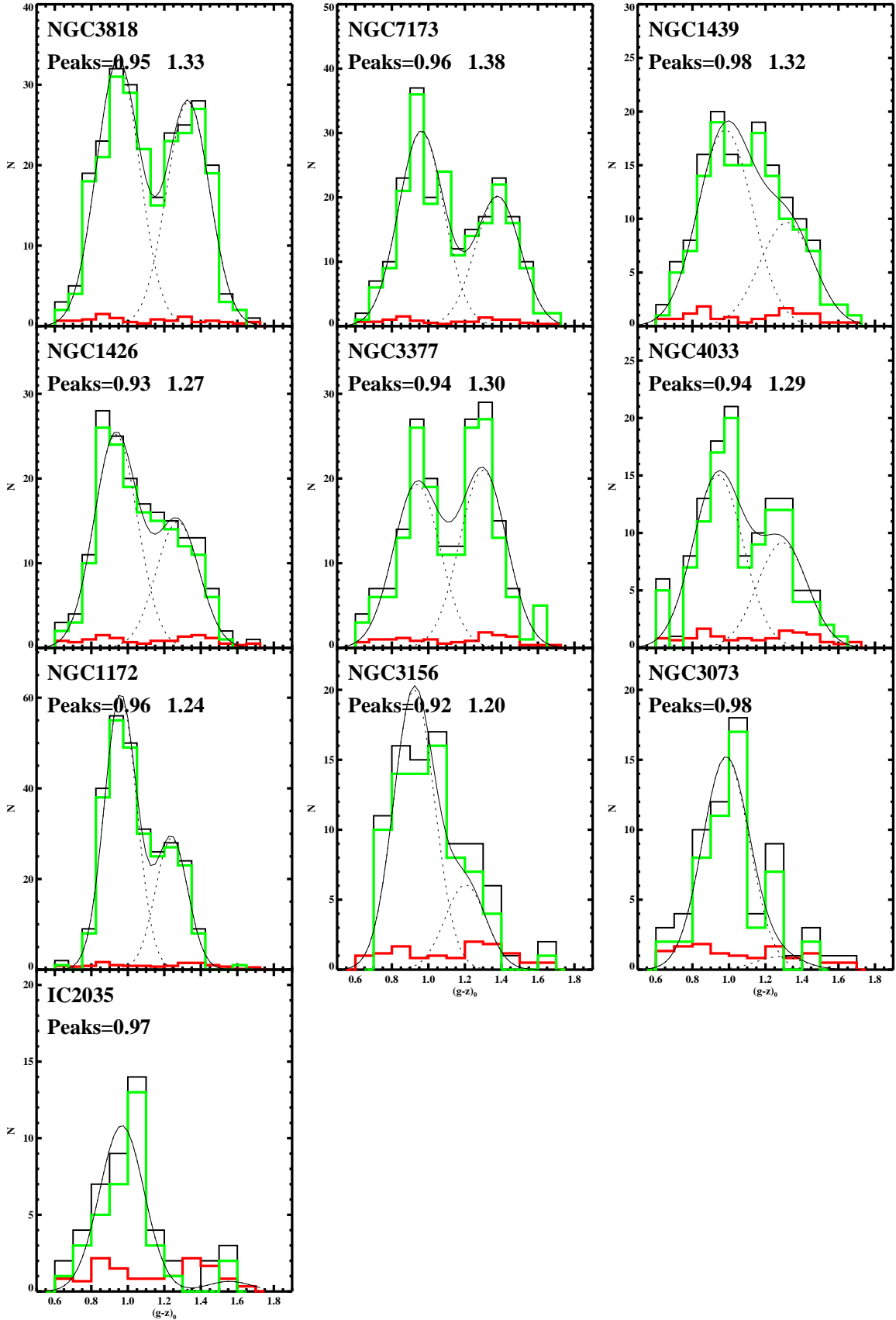}
\vspace{5mm}
       \caption[Colour distributions of GC candidates]{Colour distributions of GC candidates. The black histograms are
       colour distributions of the raw data. The red histograms are contamination estimates.
       The green histograms are after correcting for contamination. The KMM
       results for sub-populations are drawn with dashed lines, and the solid
       lines are the sum of the two sub-populations.}
     \label{fig:col_dst}     
\end{figure*}

To construct the GC colour distributions, the sample of GC candidates was refined by applying the colour cut,  $0.6 <(g-z)_0 < 1.7$, and the magnitude cut, $M_g >
-12.0$, and $g < 25.5$, where the completeness is $\sim 80\%$. These colour and magnitude selections are indicated by the dotted boxes in Figure \ref{fig:cmd}. In Figure  \ref{fig:col_dst}, the raw distribution is drawn with a black histogram. The red histogram represents the expected contamination level mentioned in the above section. For each colour bin, the number of contaminating objects was removed from the raw distribution; the green histogram represents the contamination-corrected colour distribution. The panels in Figure \ref{fig:col_dst} are placed in the same order as in Figure \ref{fig:cmd}.

\begin{table*}

\begin{center}
 \begin{minipage}{135mm}
% \begin{minipage}{110mm}
  \caption{The properties of globular cluster systems: colour distributions\label{tab:kmm}}
  %\vspace{9pt}
  \begin{tabular}{lccccccccc}
  \hline
Host galaxy & $N_{GC}$ & Blue Peak & Red Peak & $\sigma_{sub}$ & $f_{red}$ & $\langle P \rangle $ & $f(P<0.05)$ & Bimod. & $\langle (g-z)_0\rangle$  \\
\hline
\hline
(a) & \multicolumn{9}{c}{$0.6<(g-z)_0<1.7$}\\
\hline
NGC 3818 & 240 & 0.95 $\pm$ 0.01 & 1.33 $\pm$ 0.02 & 0.12 & 0.46 & 0.00 $\pm$ 0.00 & 1.00 & S & 1.12 $\pm$ 0.01 \\
NGC 7173 & 208 & 0.96 $\pm$ 0.01 & 1.38 $\pm$ 0.02 & 0.12 & 0.40 & 0.00 $\pm$ 0.00 & 1.00 & S & 1.13 $\pm$ 0.02 \\
NGC 1439 & 139 & 0.98 $\pm$ 0.03 & 1.32 $\pm$ 0.04 & 0.15 & 0.35 & 0.02 $\pm$ 0.09 & 0.63 & L & 1.10 $\pm$ 0.02 \\
NGC 1426 & 159 & 0.93 $\pm$ 0.02 & 1.27 $\pm$ 0.02 & 0.12 & 0.37 & 0.00 $\pm$ 0.00 & 0.99 & S & 1.06 $\pm$ 0.02 \\
NGC 3377 & 173 & 0.94 $\pm$ 0.02 & 1.30 $\pm$ 0.02 & 0.13 & 0.52 & 0.00 $\pm$ 0.01 & 0.96 & S & 1.13 $\pm$ 0.02 \\
NGC 4033 & 111 & 0.94 $\pm$ 0.03 & 1.29 $\pm$ 0.03 & 0.14 & 0.38 & 0.03 $\pm$ 0.16 & 0.56 & L & 1.07 $\pm$ 0.02 \\
NGC 1172 & 265 & 0.96 $\pm$ 0.01 & 1.24 $\pm$ 0.02 & 0.09 & 0.32 & 0.00 $\pm$ 0.00 & 1.00 & S & 1.05 $\pm$ 0.01 \\
NGC 3156 &  74 & 0.95 $\pm$ 0.05 & 1.29 $\pm$ 0.20 & 0.13 & 0.16 & 0.03 $\pm$ 0.09 & 0.57 & L & 1.01 $\pm$ 0.02 \\
NGC 3073 &  52 & 0.98 $\pm$ 0.03 & ... & 0.13 & 0.06 & 0.28 $\pm$ 0.42 & 0.16 & U & 1.02 $\pm$ 0.02 \\
IC 2035 &  35 & 0.97 $\pm$ 0.03 & ... & 0.12 & 0.06 & 0.00 $\pm$ 0.01 & 0.83 & U & 1.00 $\pm$ 0.03 \\
\hline
\hline
(b) & \multicolumn{9}{c}{$0.7<(g-z)_0<1.6$}\\
\hline
NGC 3818 & 236 & 0.95 $\pm$ 0.01 & 1.33 $\pm$ 0.01 & 0.11 & 0.46 & 0.00 $\pm$ 0.00 & 1.00 & S & 1.12 $\pm$ 0.01 \\
NGC 7173 & 201 & 0.97 $\pm$ 0.01 & 1.37 $\pm$ 0.02 & 0.11 & 0.40 & 0.00 $\pm$ 0.00 & 1.00 & S & 1.14 $\pm$ 0.02 \\
NGC 1439 & 134 & 0.98 $\pm$ 0.02 & 1.30 $\pm$ 0.04 & 0.13 & 0.39 & 0.01 $\pm$ 0.02 & 0.84 & L & 1.10 $\pm$ 0.02 \\
NGC 1426 & 157 & 0.94 $\pm$ 0.02 & 1.28 $\pm$ 0.02 & 0.11 & 0.38 & 0.00 $\pm$ 0.00 & 1.00 & S & 1.07 $\pm$ 0.02 \\
NGC 3377 & 165 & 0.95 $\pm$ 0.01 & 1.30 $\pm$ 0.01 & 0.11 & 0.53 & 0.00 $\pm$ 0.00 & 1.00 & S & 1.13 $\pm$ 0.02 \\
NGC 4033 & 105 & 0.95 $\pm$ 0.01 & 1.30 $\pm$ 0.02 & 0.11 & 0.40 & 0.00 $\pm$ 0.00 & 0.98 & L & 1.09 $\pm$ 0.02 \\
NGC 1172 & 264 & 0.96 $\pm$ 0.01 & 1.24 $\pm$ 0.01 & 0.09 & 0.33 & 0.00 $\pm$ 0.00 & 1.00 & S & 1.05 $\pm$ 0.01 \\
NGC 3156 &  73 & 0.92 $\pm$ 0.03 & 1.20 $\pm$ 0.05 & 0.11 & 0.23 & 0.09 $\pm$ 0.15 & 0.40 & L & 1.00 $\pm$ 0.02 \\
NGC 3073 &  50 & 0.99 $\pm$ 0.03 & ... & 0.11 & 0.16 & 0.08 $\pm$ 0.27 & 0.42 & U & 1.03 $\pm$ 0.02 \\
IC 2035 &  34 & 0.98 $\pm$ 0.02 & ... & 0.11 & 0.06 & 0.00 $\pm$ 0.00 & 0.88 & U & 1.01 $\pm$ 0.03 \\
\hline

\end{tabular}
\medskip
 
\end{minipage}
\end{center}
\end{table*}
%}}

In order to statistically test the significance of any colour bimodality, the Kaye's mixture model (KMM) test \citep{ash94} was employed. The KMM test uses likelihood ratio test statistics to estimate the probability (P-value) that two distinct Gaussians with the same dispersion are a better fit to the observational data than a single Gaussian. After running the KMM test for 100 bootstrap resamples of the contamination corrected colours, the peaks of the colour histograms, the $\sigma$ of the sub-populations, and the red population fraction were estimated. Also estimated were the P-value and the fraction of P-values that are less than 0.05 (Table \ref{tab:kmm}(a)); the representative values are median values of the 100 KMM outputs and the uncertainties are half a width within which 68\% of the data are contained relative to the median. Since the KMM outputs (especially the P-values) have a somewhat skewed distribution, the median is a more robust estimation than the mean. The statistical interpretation of the P-value is that a distribution with a P-value of 0.05 favors bimodality rather than unimodality with a 95\% confidence. In Figure  \ref{fig:col_dst}, the results of the KMM tests for each subgroup are plotted with dashed lines, and the sum of the two groups is plotted with solid lines. The significance of colour bimodality is decided on the basis of $\langle P \rangle$ and the fraction of $P<0.05$. If $\langle P \rangle < 0.05$ and $f_{(P<0.05)} > 0.90$, a distribution is interpreted as a strong bimodal distribution and for $0.05 < \langle P \rangle < 0.10$ or $0.50 < f_{(P<0.05)} < 0.90$, a distribution is likely bimodal; otherwise it is unimodal. Despite the small P-value for IC 2035, this galaxy is regarded as unimodal because of the insignificant number of red clusters. The significance of bimodality is listed as S(trong), L(ikely), or U(nimodal) in Table~\ref{tab:kmm}.

\begin{figure*}
\includegraphics[width=170mm]{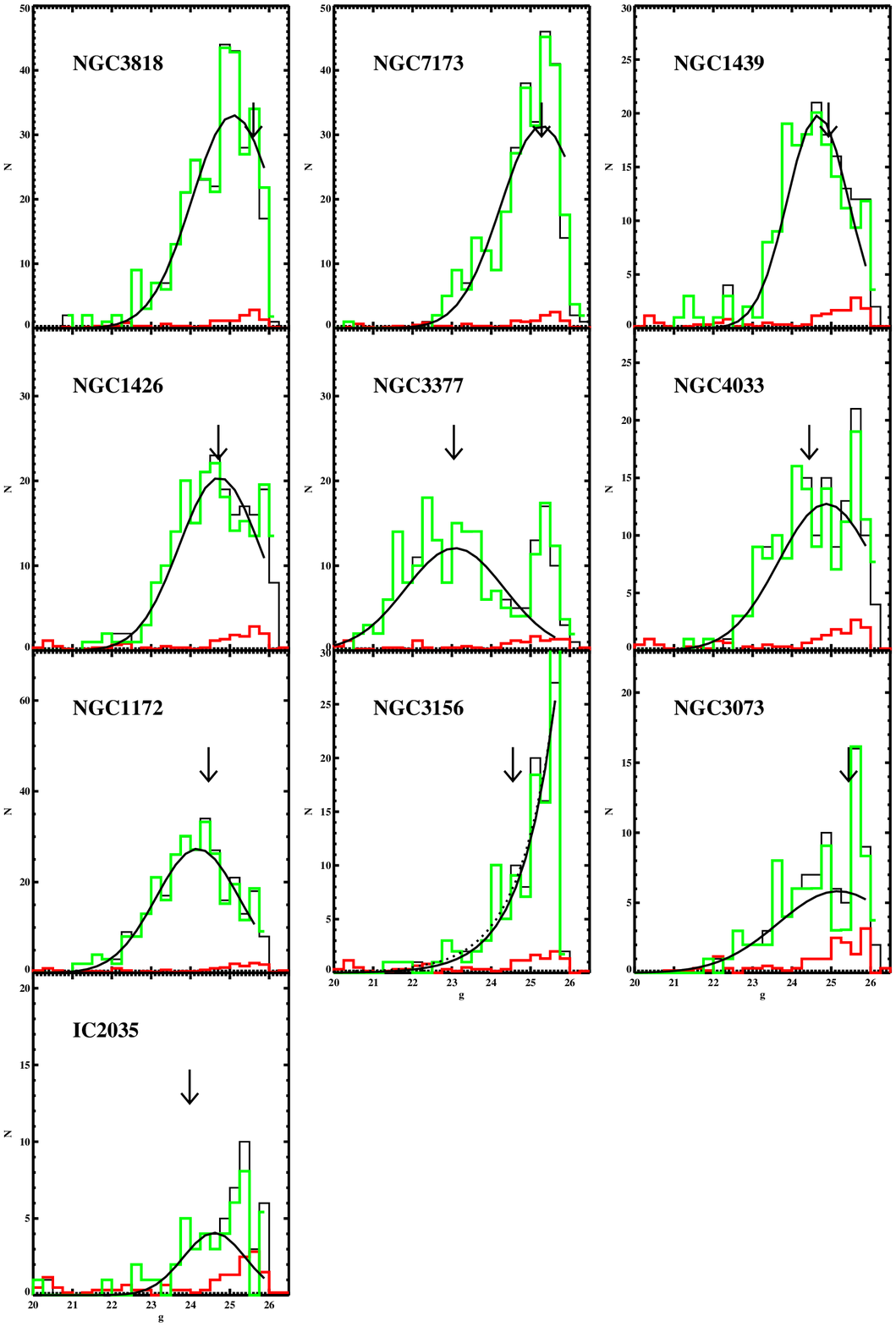}
\vspace{3mm}
       \caption[Luminosity functions of GCs in the $g$ band]{Luminosity functions of GCs in the $g$ band. The black histograms are luminosity functions before correction. The red histograms are contamination estimates, and the green ones are  luminosity functions corrected for background contamination and completeness. The solid curves are the best fit Gaussian luminosity functions, and the arrows indicate the expected turn-over position for each galaxy based on the GCLF parameters from \citep{jor07b}.}
     \label{fig:gclf_g}     
\end{figure*}

\begin{figure*}
\includegraphics[width=170mm]{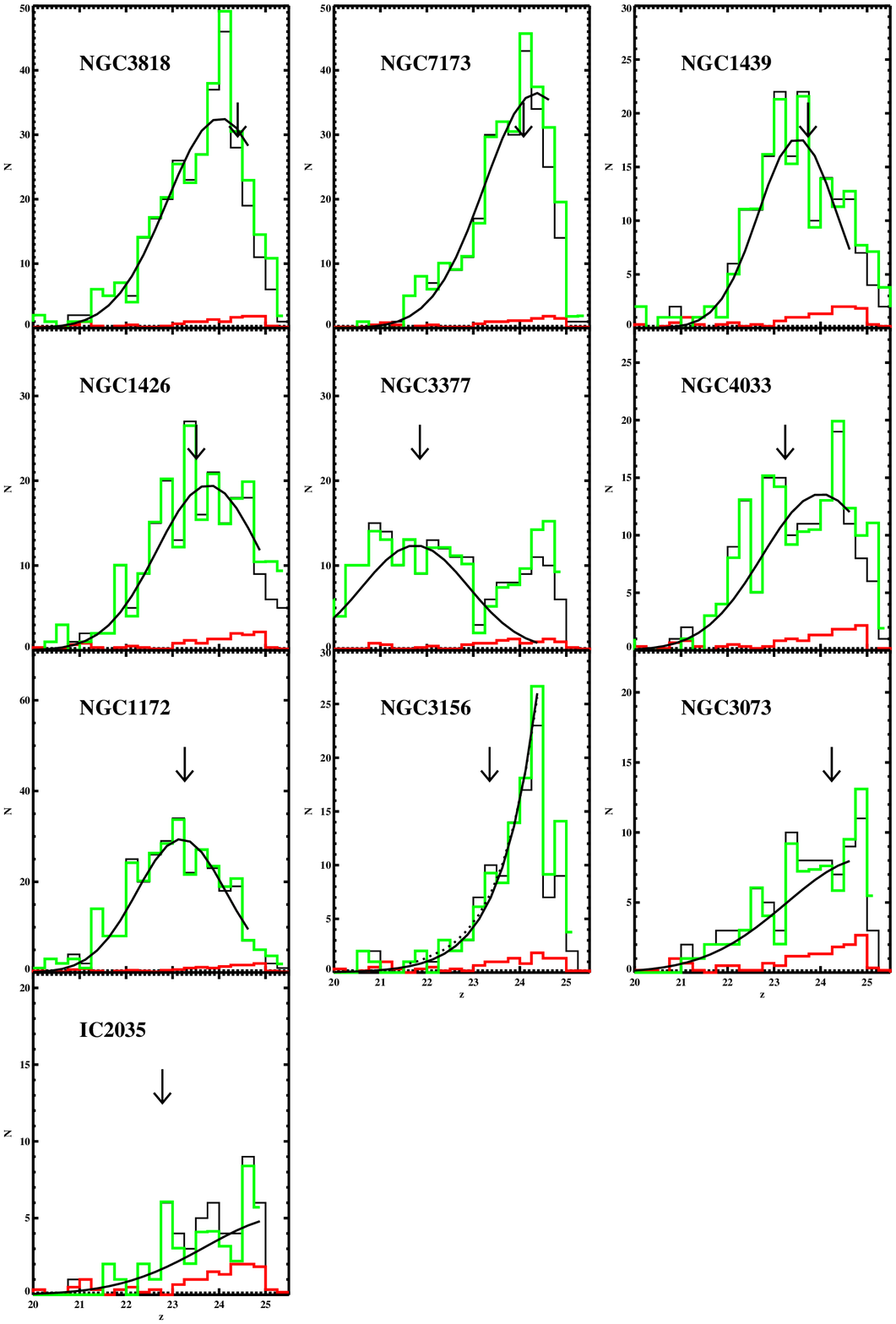}
\vspace{3mm}
       \caption[Luminosity functions of GCs in the $z$ band]{Luminosity functions of GCs in the $z$ band. The histogram colours and fitted curves have the same meaning as for the $g$ band luminosity functions in Figure \ref{fig:gclf_g}.}
     \label{fig:gclf_z}     
\end{figure*}

The ability of the KMM test to detect bimodality depends on the sample number and the normalized separation between the two peaks (see \citealt{ash94}).
A larger sample size and a larger separation between two subcomponents will increase the chance of detecting bimodality. Furthermore, the extended tails of the distribution can also affect the results. Therefore, an alternative GC sample with a narrower colour range of $0.7<(g-z)_0<1.6$ was tested. The corresponding results are listed in Table~\ref{tab:kmm}(b). Clipping the tails of the distributions was found to decrease the P-value. In other words, without the extended tails, the KMM test is more likely to detect bimodality. This is not surprising because fitting extended tails results in a larger sigma for each subgroup, thus making it harder to separate two subgroups. However, in this experiment, the values of most of the fitted parameters do not vary significantly and remain within the uncertainty that was initially estimated. There is one case where a fitted parameter, the red peak colour, has been changed significantly after clipping the tails. This scenario occurs in NGC 3156, where one isolated and very red GC (see middle-bottom panel in Figure \ref{fig:col_dst}) makes the red peak significantly redder compared to the value when it is removed. This is potentially a background galaxy contaminant because, unlike other red GCs, it lies at a large projected distance from the galaxy center. For further analysis, the values from Table~\ref{tab:kmm}(a) were adopted in this study except for NGC 3156, where the value from Table~\ref{tab:kmm}(b) is used. The KMM results are overplotted in Figure~\ref{fig:col_dst}.

It should be noted that colour bimodality in GCs is more common in the more luminous galaxies. The red population becomes weaker and moves to bluer colours with decreasing galaxy luminosity. Eventually, only a blue peak appears in the faintest galaxies. The richness of the GCs also decreases with the host galaxy luminosity. None of our GC colour distributions have a single broad peak located between the normal blue and red peaks. This is in contrast to the results of some studies on early-type galaxies in other environments (e.g. \citealt{lar01}; \citealt{kun01a}; \citealt{pen06a}). The behavior of the GC colour with host galaxy luminosity will be discussed in more detail below.

\subsection{Luminosity functions}
\label{sec:gclf_low}

\begin{table*}
\begin{center}
% \begin{minipage}{126mm}
% \begin{minipage}{130mm}
  \caption{Gaussian fitted parameters of luminosity functions and specific frequencies \label{tab:gclf}}
 % \vspace{9pt}
  \begin{tabular}{lccccc}
\hline
 Host galaxy & $m_{to,g}$ & $\sigma_g$ & $m_{to,z}$ & $\sigma_z $ & $S_N$\\
\hline
NGC 3818 & 25.09 $\pm$ 0.13 & 1.05 $\pm$ 0.09 & 24.02 $\pm$ 0.17 & 1.14 $\pm$ 0.11 & 1.36 $\pm$ 0.12 \\
NGC 7173 & 25.28 $\pm$ 0.15 & 1.04 $\pm$ 0.11 & 24.36 $\pm$ 0.24 & 1.12 $\pm$ 0.14 & 1.63 $\pm$ 0.06 \\
NGC 1439 & 24.66 $\pm$ 0.08 & 0.77 $\pm$ 0.07 & 23.51 $\pm$ 0.10 & 0.84 $\pm$ 0.08 & 0.91 $\pm$ 0.10 \\
NGC 1426 & 24.73 $\pm$ 0.12 & 1.02 $\pm$ 0.09 & 23.77 $\pm$ 0.14 & 1.11 $\pm$ 0.11 & 1.67 $\pm$ 0.21 \\
NGC 3377 & 23.08 $\pm$ 0.12 & 1.25 $\pm$ 0.11 & 21.75 $\pm$ 0.11 & 1.14 $\pm$ 0.09 & 2.35 $\pm$ 0.49\\
NGC 4033 & 24.88 $\pm$ 0.24 & 1.22 $\pm$ 0.17 & 24.01 $\pm$ 0.33 & 1.26 $\pm$ 0.20 & 1.91 $\pm$ 0.19\\
NGC 1172 & 24.18 $\pm$ 0.09 & 1.06 $\pm$ 0.07 & 23.19 $\pm$ 0.07 & 0.96 $\pm$ 0.06 & 9.18 $\pm$ 4.41\\
NGC 3156 & \ldots & \ldots & \ldots & \ldots & 0.62 $\pm$ 0.02\\
NGC 3073 & 25.17 $\pm$ 0.67 & 1.53 $\pm$ 0.44 & \ldots & \ldots & 3.48 $\pm$ 1.76\\
IC 2035 & 24.60 $\pm$ 0.19 & 0.79 $\pm$ 0.17 & \ldots & \ldots & 0.91$\pm$ 0.15\\
\hline
\end{tabular}
%\medskip

%\end{minipage}
\end{center}

\end{table*}

\begin{figure*}
\begin{center}
\includegraphics[width=130mm]{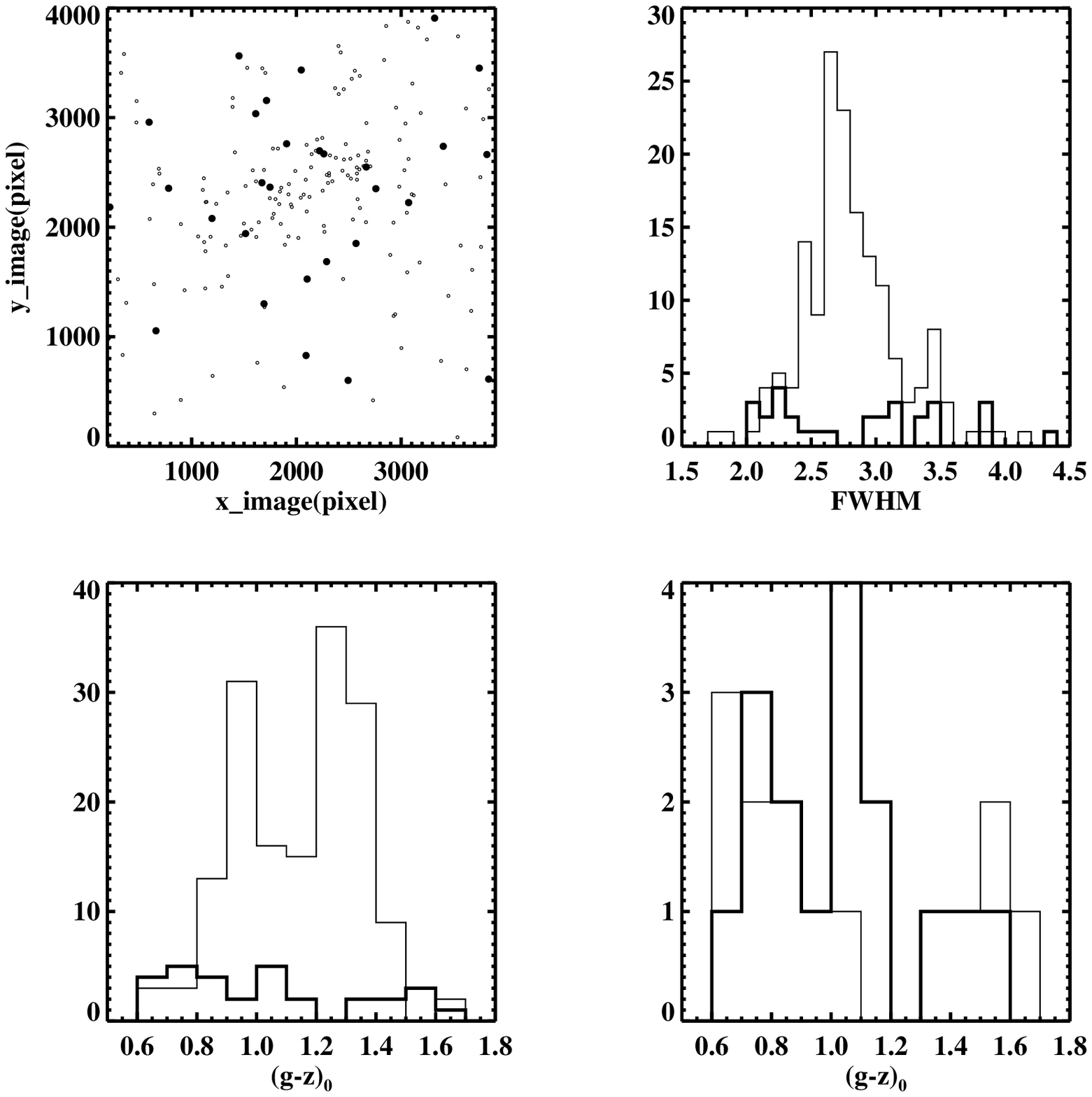}
\vspace{3mm}
       \caption[Properties of faint GCs near $g\sim25.5$ mag in NGC 3377]{Properties of faint GCs near $g \sim 25.5$ mag in NGC 3377. The top-left panel shows the spatial distribution of bright GCs ($g< 25$ mag, small open circles) and faint GCs ($g> 25$ mag, large closed circles). The top-right panel shows the histograms of FWHM of the bright (thin histogram) and faint (thick histogram) GCs. The bottom-left panel shows colour distributions of the bright (thin histogram) and faint (thick histogram) GCs. The bottom-right panel shows colour distributions of the faint GCs with larger $FWHM> 2.8$ (thin histogram) and smaller $FWHM<2.8$ (thick histogram).}
     \label{fig:n3377star}
\end{center}     
\end{figure*}

Luminosity functions in the $g$ and $z$ bands were constructed using GCs in the colour range of $0.6 <(g-z)_0 < 1.7$ (black histograms in Figure~\ref{fig:gclf_g} and Figure~\ref{fig:gclf_z}). The contamination (red histograms) estimated previously was subtracted from the raw distribution and incompleteness corrections were applied. The corrected luminosity functions (green histograms) were fitted with a Gaussian function up to the 50\% completeness limit. The fitted parameters (peak magnitude and standard deviation) and their uncertainties are listed in Table~\ref{tab:gclf}; the missing data correspond to parameters with a large uncertainty caused by the small number of GC samples and an excess of faint GCs. For the luminous galaxies, Gaussian functions aptly represent the luminosity functions of the GC systems. The arrows in Figures~\ref{fig:gclf_g} and \ref{fig:gclf_z} indicate the expected turn-over magnitudes based on a peak absolute magnitude of the GC luminosity function (GCLF) of $\mu_g=-7.2$ and $\mu_z=-8.4$ as given by \citet{jor07b}.

% For NGC 3377, the
%luminosity function is deviated from the Gaussian function in the faint end ($g\sim25.5mag),
%possessing a large number of faint GCs. We speculated a possibility of 

As can be seen in Figure \ref{fig:gclf_g}, there is an excess of faint GCs around $g\sim25.5$ mag in the GCLF of NGC 3377. These faint GCs deviate significantly from the fitted Gaussian function. This deviation is also found at $z\sim24.5$ mag in the $z$-band GCLF shown in Figure \ref{fig:gclf_z}. The properties of these objects are discussed in more detail below. 

In Figure \ref{fig:n3377star}, the spatial distribution, size, and colour of NGC 3377 GCs fainter than $g=25$ mag are plotted along with those brighter than $g=25$ mag. In the top-left panel, the faint GCs (large closed circles) appear to be uniformly distributed, unlike the bright GCs (small open circles) which are concentrated at the galaxy center. Shown in the top-right panel of Figure 8 are histograms of the FWHM returned by {\it Sextractor} for the faint and bright GCs. The size distribution of the faint GCs is also quite different from that of the bright GCs in the sense that the sizes are either smaller or larger than the medium size for all clusters. The colour distribution of the faint GCs appears to be rather flat, as shown in the bottom-left panel. Any dependence of the colour on the size of the faint GCs was examined by separating the GCs at $FWHM=2.8$ pixels. In the bottom-right panel of Figure 8, the thin histogram represents the colour distribution of faint GCs with $FWHM >2.8$, while the thick histogram represents GCs with $FWHM<2.8$. No obvious correlation between size and colour was found in the plot.

We can speculate on the possible origin of the faint objects that deviate from the normal Gaussian GCLF. \citet{pen06b} found a class of diffuse star clusters (DSCs) in the ACSVCS and investigated the nature of these objects, which are characterized by their low-luminosity, broad distribution of sizes, low surface brightness, and redder colour when compared to normal metal-rich GCs. The spatial distribution of the objects was also closely associated with the host galaxy light. However, not all of the aforementioned characteristics are found in our faint objects, which possess a wide distribution of sizes, but a significant number of blue GCs and a random spatial distribution.

Another possibility is that the faint objects may be field stars (bright giants) that belong to the parent galaxy. By comparing isochrones of various ages and metallicities (\citealt{gir06}), the locus of giant stars ($M_V$ $<$ $-5$) of a very young age ($\sim$10 Myr) was found to overlap with that of the faint objects on the colour-magnitude diagram (Figure \ref{fig:cmd}). However, it is quite unlikely that such young stars exist in significant numbers in old elliptical galaxies like NGC 3377. Furthermore, horizontal branch stars are not sufficiently luminous to account for the faint objects. \citet{har07} found no evidence of young stellar populations ($<3$ Gyr) in the halo star CM diagram of NGC 3377 from deep HST/ACS photometry, although their field of view barely overlaps that used in this study. The number density of foreground Milky Way stars in the direction of NGC 3377 is also very low ($N_{star}<10$) within the ACS field (\citealt{ro03}) and even lower in the magnitude range of the faint GCs. Visual inspection of the faint objects reveals that some are more likely to be misclassified background galaxies.  Assuming the faint objects are not GCs belonging to NGC3377 shifts the mean colour of the GC system blueward by ($\Delta(g-z)_{0}\sim0.015$) compared to the value including all the objects. Note that the mean colours in Table \ref{tab:kmm} have been estimated after a faint magnitude cut ($g=25.5$ mag) so that any colour change by faint objects is negligible.

%mean colour for all gcs1.124
%mean colour excluding faint 1.141
\subsection{Host galaxy properties  vs. globular cluster colours}

\begin{figure} 
\begin{center}
\includegraphics[width=85mm]{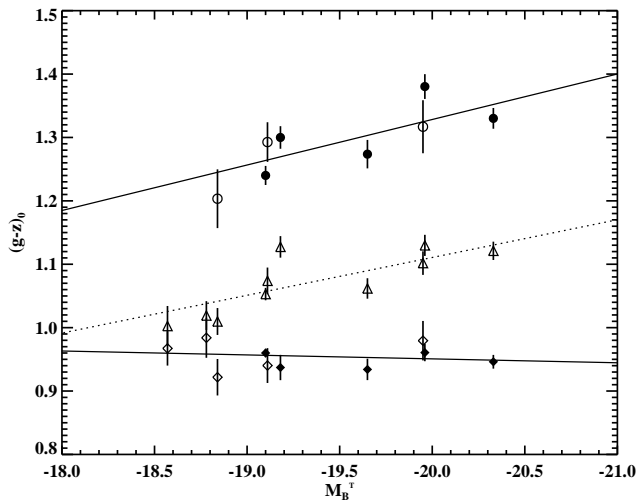} 
\caption[A plot of GC colours against host galaxy luminosity]{A plot of 
GC colours against host galaxy luminosity. The diamond and circle symbols
represent the blue and red peaks respectively. Strong bimodality is indicated
by a filled symbol,and marginal or zero bimodality by an open symbol. The mean
colours of all GCs within a galaxy are plotted with an open triangle. The two solid lines are the least squares best fits
to the blue and red peaks. The dotted line is fitted to the mean colour.} 
\label{fig:mb_col} 
\end{center}
\end{figure}

\begin{figure}
\begin{center}
\includegraphics[width=87mm]{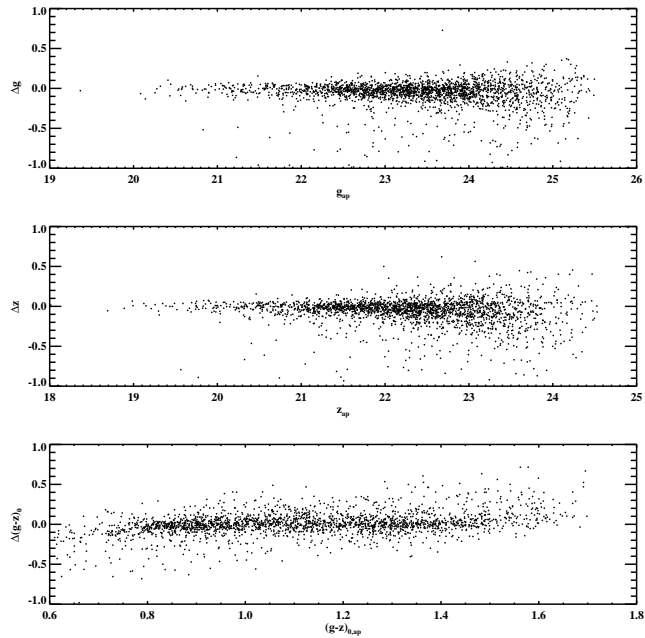} 
\caption[]{Differences in $g$ and $z$ magnitudes and $g-z$ colour for GCs in 24 ACSVCS galaxies between our aperture photometry and {\it KINGPHOT} from \citet{jor09}. The measurement of our aperture photometry is on the x-axes and offsets between the two measurements are on the y-axes. From top to bottom the offsets plotted are: $\Delta g=g_{ap}-g_{King}$, $\Delta z=z_{ap}-z_{King}$ and $\Delta(g-z)_{0}=(g-z)_{0,ap}-(g-z)_{0,King}$.} 
\label{fig:com_mag} 
\end{center}
\end{figure}

It is well known that there is a strong correlation between mean colours/red peaks and host galaxy luminosities (e.g. \citealt{lar01}; \citealt{kun01a}). However, few studies had been performed to support a correlation between the blue peak and the host galaxy luminosity until the recent ACSVCS data found such a relationship with their large dynamic range of galaxy luminosity \citep{pen06a}.

In Figure~\ref{fig:mb_col}, the peak colours of the sub-populations and the mean GC colours are plotted against the host galaxy luminosity for our sample. Each relationship is fitted to a straight line using weighted chi-squre minimization. It is clear that strong dependencies of the red peak and the mean GC colour on the host galaxy luminosity do exist in our field galaxies with almost identical slopes. On the other hand, the blue peak is almost independent of the galaxy luminosity. In fact, the fitted slope exhibits a weak anti-correlation, the statistical significance of which is within 1$\sigma$ of a zero slope. From the attained results alone, it is uncertain whether the absence of a blue peak gradient is due to the narrow dynamic range of the luminosity in our field galaxy sample or an intrinsic effect across a wide range of luminosity. The obtained results clearly show that the observed trends in the GC colours with host galaxy luminosity are a consequence of the gradual change in the fraction and colour of the red GCs and not a systematic change in the colours of all GCs within a galaxy.

\subsection{Comparison with the ACS Virgo Cluster Survey}
\label{sec:com_vcs}

\begin{figure}
\begin{center}
\includegraphics[width=87mm]{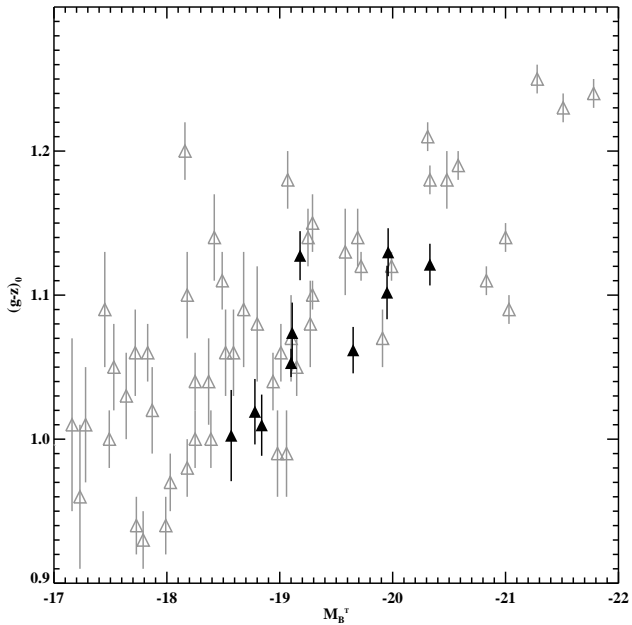} 
\caption[Comparison of mean colours for GCs in our galaxy sample with those from the ACS Virgo Cluster Survey]{Comparison of mean colours for GCs in our galaxy sample with those from the ACS Virgo Cluster Survey (ACSVCS). Our results  are plotted with black filled triangles, while ACSVCS results from PJC06 are plotted with grey open triangles. The mean colours of our field galaxy sample appear to be situated at the bluer end of the colour range for a given galaxy luminosity.} 
\label{fig:mb_colmean_com} 
\end{center}
\end{figure}

\begin{figure}
\begin{center}
\includegraphics[width=87mm]{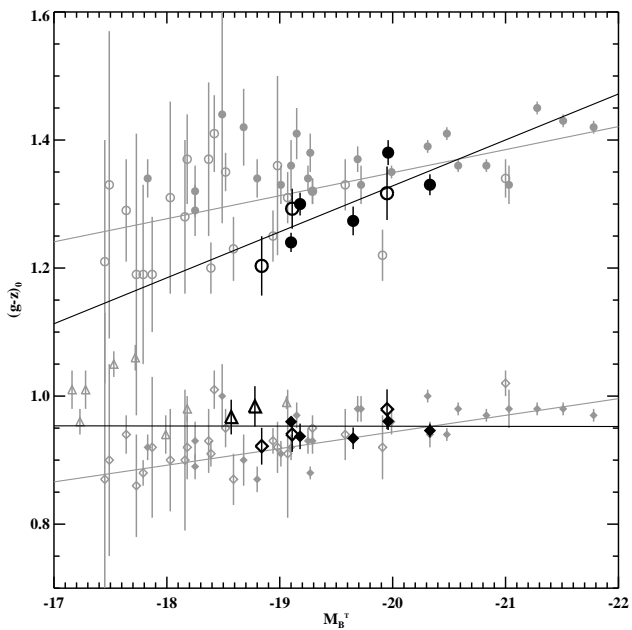} 
\caption[Comparison of colour peaks with the ACSVCS]{Comparison of colour peaks with the ACSVCS. Symbols are identical with Figure \ref{fig:mb_col}. Black symbols are from our results, while grey symbols are from the ACSVCS. The dark sold lines are a linear fit to each sub-population from our results, whereas the grey solid lines are linear fits by PJC06 to the ACSVCS.} 
\label{fig:mb_colpeak_com} 
\end{center}
\end{figure}

\begin{figure}
\begin{center}
\includegraphics[width=86mm]{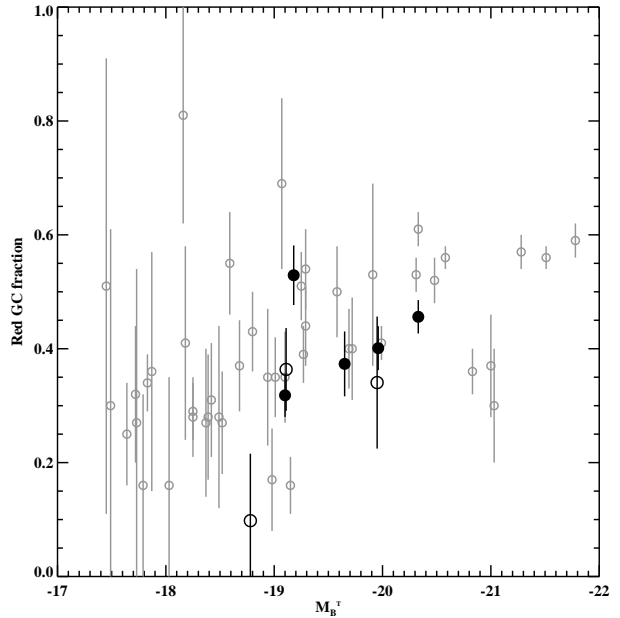} 
\caption[Comparison of the fraction of red GCs with the ACSVCS]{Comparison of fraction of red GCs  with the ACSVCS. The black filled and open circles represent strong and likely bimodality from this work respectively, while open circles are from the ACSVCS. No corrections have been made for any aperture bias effects between the two samples (see text for details).} 
\label{fig:mb_rf_com} 
\end{center}
\end{figure}

The ACS Virgo Cluster Survey (ACSVCS) was a large program used to image 100 early-type galaxies in the Virgo cluster with the ACS in two filters ($g$ and $z$ bands) \citep{cot04}. The main scientific goals were to study the properties of GC systems in these galaxies (e.g., PJC06, \citealt{jor06}), analyze their central structures \citep{fer06}, and obtain accurate surface brightness fluctuation distances (\citealt{mei07}; see also \citealt{bla09}).

The observation conditions and data reduction procedures of the ACSVCS are somewhat different from those adopted in this study. The ACSVCS used 100 orbits of the HST, allocating only one orbit for each galaxy. Each galaxy of the ACSVCS had an exposure time of 750 s in F475W and 1210 s in the F850LP filter, while our exposure times are almost twice as long (typically 1300 s in F457W and 3000 s in F850LP). The data reduction procedures of the ACSVCS are described in detail by \citet{jor04}. In brief, the exposures were split into two or three sub-exposures, without repositioning, so as to remove cosmic rays. These sub-exposures were then reduced and combined using the standard ACS pipeline. Model galaxies were created using the {\it ELLIPROF} program \citep{ton97} and subtracted from the original images. The {\it KINGPHOT} program developed by Jordan et al. (2005) was used for photometry of the GCs and to measure their sizes. {\it KINGPHOT} fits King models convolved with a given point spread function in each filter to individual GC candidates and finds the best fit parameters of the King model. Total magnitudes for the GCs were calculated by integrating these best-fit convolved King models. All GCs were selected based on their magnitude ($g\geq19.1$ or $z\geq18.0$) and mean elongation ($\langle e \rangle \leq2$, where $\langle e \rangle \equiv a/b$) in the two filters. The main results regarding the colour distributions of the ACSVCS GC systems were published by PJC06 and are compared with the results of this study below.

Before directly comparing results from the ACSVCS with ours, it is necessary to check consistency of photometry, because the applied photometry methods are different; aperture photometry has been used in this work, while the ACSVCS used fitted King models. We applied our photometry procedure to 24 galaxies of the ACSVCS within our galaxy luminosity range and directly compared magnitudes and colours of GCs we measured with those in the GC catalogue from \citet{jor09}. Their catalogue lists two kinds of photometry values, PSF convolved King model and aperture photometry. Since PJC06 used the former for their analysis, it was taken here for comparison. Figure \ref{fig:com_mag} shows differences of $g$ and $z$ magnitudes ($\Delta g=g_{ap}-g_{King}$,$\Delta z=z_{ap}-z_{King}$) and colour ($\Delta(g-z)_{0}=(g-z)_{0,ap}-(g-z)_{0,King}$) for all the GCs matched in 24 ACSVCS galaxies. It is found that mean magnitude offsets in $g$ and $z$ band are $\langle\Delta g\rangle\sim0.025$ and $\langle\Delta z\rangle\sim0.024$ respectively, revealing that our photometry appears to slightly overestimate total flux of GCs. However, no offsets in mean colour  between the two methods are seen, $\langle\Delta(g-z)_{0}\rangle<0.01$. This test verifies that there is no systematic colour offset between the two methods so that direct comparison of GC colours between PCJ06's results and our results can be justified.

In Figure \ref{fig:mb_colmean_com}, the mean colours for the 10 GC systems in our sample are compared with the results from PJC06. At a given host galaxy luminosity, our mean colours are generally located at the bluer end of the mean colour distribution of the ACSVCS galaxies. However, unlike the other GC systems in our sample, the mean colour of the NGC 3377 GC system lies in the middle of the range of ACSVCS mean colours. Because the distribution of red GCs is more centrally concentrated and NGC 3377 is the nearest galaxy in our sample, it is possible that the GCs detected within our ACS field are biased toward red GCs. To test whether this selection bias is large enough to shift the mean colour, the GCs in other galaxies were resampled with a smaller field of view and their mean colours were recalculated. For this test NGC 4033 and NGC 1172, which have a similar absolute magnitude but are as twice as distant as NGC 3377, were chosen. The mean colour of the GCs in these two galaxies within a 50$\arcsec$ radius from their galaxy centers is $\sim0.03$ mag redder than our original estimations for the full spatial coverage. Noting that the full field of view of the ACS is $202\arcsec\times202\arcsec$ so that a circle with a 50$\arcsec$ radius covers almost half of the original aperture, the intrinsic mean colour of the GCs in NGC 3377 could be about $\sim0.03$ mag bluer than our original estimation. However, this is a relatively small correction when compared to the wide range of mean colours observed in the ACSVCS. Note that the error bar of NGC 3377 plotted in Figure \ref{fig:mb_colmean_com} is also $\pm0.02$ mag.

PJC06 analyzed their colour distributions with two different methods. One method separates the two sub-populations in each host galaxy and tests for bimodality using the KMM routine. The other method coadds the colour distributions of GCs in bins of the host galaxy luminosity and separates the two sub-populations by nonparametric decomposition. The results from the first method of PJC06 were compared with our results because our colour distributions were also analyzed with the KMM method and the number of our sample galaxies is not large enough or wide enough to bin by galaxy luminosity. A comparison of our results and those of PJC06 with respect to the colour peaks of the two sub-populations is shown in Figure \ref{fig:mb_colpeak_com}. A comparison of the fractions of red population GCs detected was also plotted; the results are shown in Figure \ref{fig:mb_rf_com}. In Figure \ref{fig:mb_rf_com}, only galaxies with a bimodal colour distribution are plotted.

\begin{figure*}
\begin{center}
\includegraphics[width=184mm]{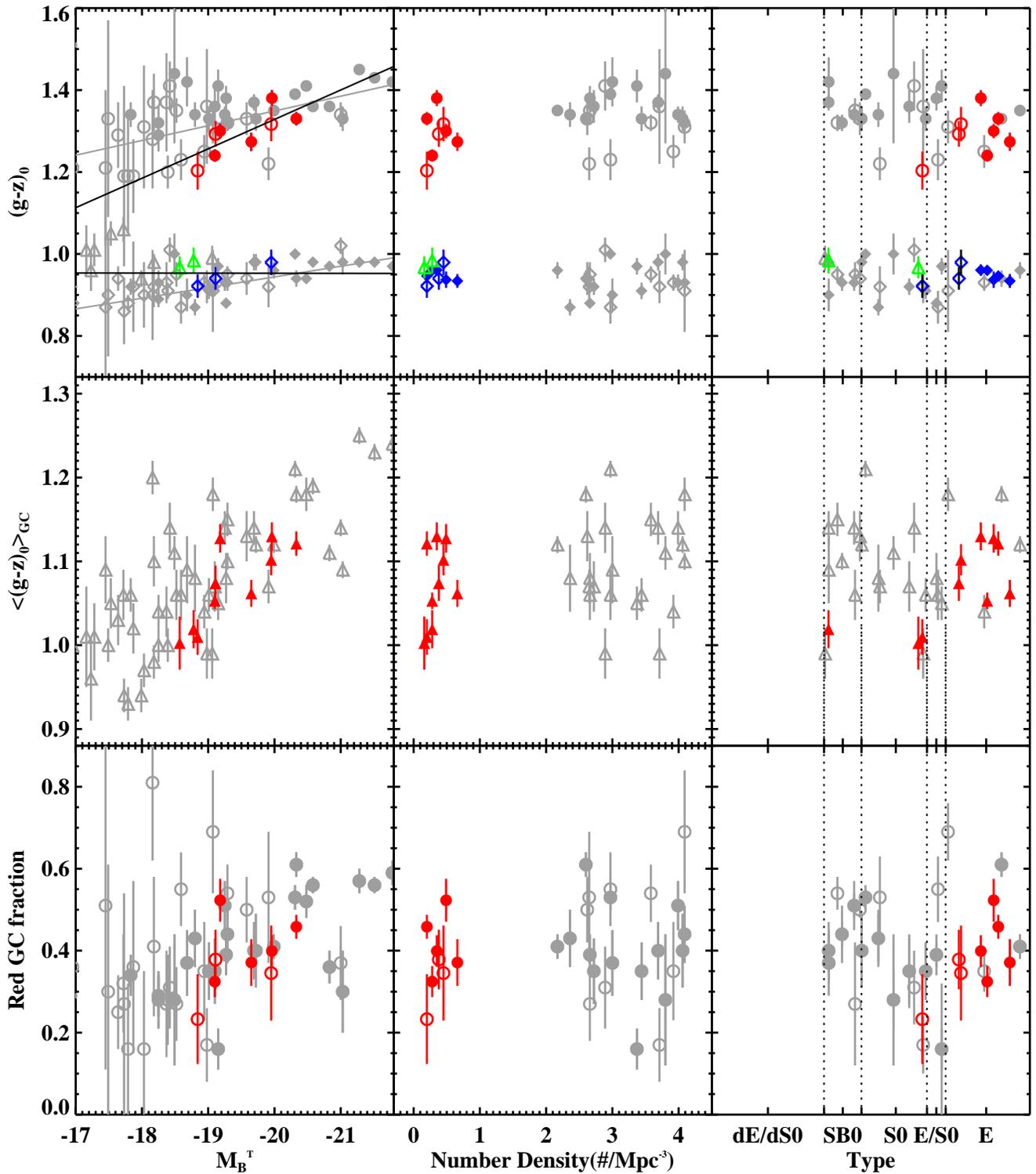} 
\caption[Comparison of the global properties of GC systems in our sample of low-luminosity field E/S0s with those of the ACSVCS]{Comparison of the global properties of GC systems in our sample of low-luminosity field E/S0s with those of the ACSVCS. The first, second, and third columns show plots versus the absolute magnitude, the local density of galaxies from Tully (1988), and the morphological type of the host galaxy, respectively. The rows show the colour of the blue and red peaks, mean colour, and the fraction of red GCs. The coloured symbols represent our data, while the grey symbols are from the ACSVCS. For the second and third columns, only galaxies from the ACSVCS within the luminosity range of our sample are plotted. For the first and third rows, closed circles show the GCs in which colour bimodality is strongly detected using the KMM test, while open symbols show GCs with weak colour bimodality. In the top-left panel, black lines are the best linear fit to our data and the grey lines are the best-fit to the ACSVCS data.
 } 
\label{fig:gal_gc_com} 
\end{center}
\end{figure*}

Overall trends for the colour distributions with respect to the host galaxy luminosity in both samples appear to be similar. As evident in Figures \ref{fig:mb_colpeak_com} and \ref{fig:mb_rf_com}, the probability of bimodality decreases with decreasing host galaxy luminosity as the red population becomes weaker. There is a strong correlation between the peak of the red sub-population and the host galaxy luminosity. Moreover, the fraction of red GCs with respect to the entire population appears to increase with host galaxy luminosity.

Some differences do exist between the results of this study and those of PJC06. In Figure \ref{fig:mb_colpeak_com}, our slope for the red peak against galaxy absolute magnitude appears to be steeper than that of the ACSVCS. In other words, the bimodal colour distributions in our sample disappear more quickly as galaxies become fainter. Furthermore, we found no correlation between the blue peaks and the galaxy absolute magnitude, while PJC06 found a weak correlation (although it was not as steep as for the red peaks). Our fraction of red GCs was also found to be slightly lower than that of the ACSVCS at a given galaxy luminosity. The mean colours of the GCs in our sample are slightly bluer than those of the ACSVCS and the red and blue peak positions in our sample are almost identical to those in the ACSVCS. This suggests that the difference in the mean colours is due to differences in the relative fraction of the red population and not an overall shifting of colours. In the middle bottom panel of Figure \ref{fig:gal_gc_com} it appears as if the ACSVCS galaxies have more red objects. To test whether this is real or not, a K-S test was performed, but its results give a low significance level to any difference.

\begin{figure*}
  %\vspace{9pt}

  \centerline{\hbox{ \hspace{0.0in} 
    \includegraphics[width=84mm]{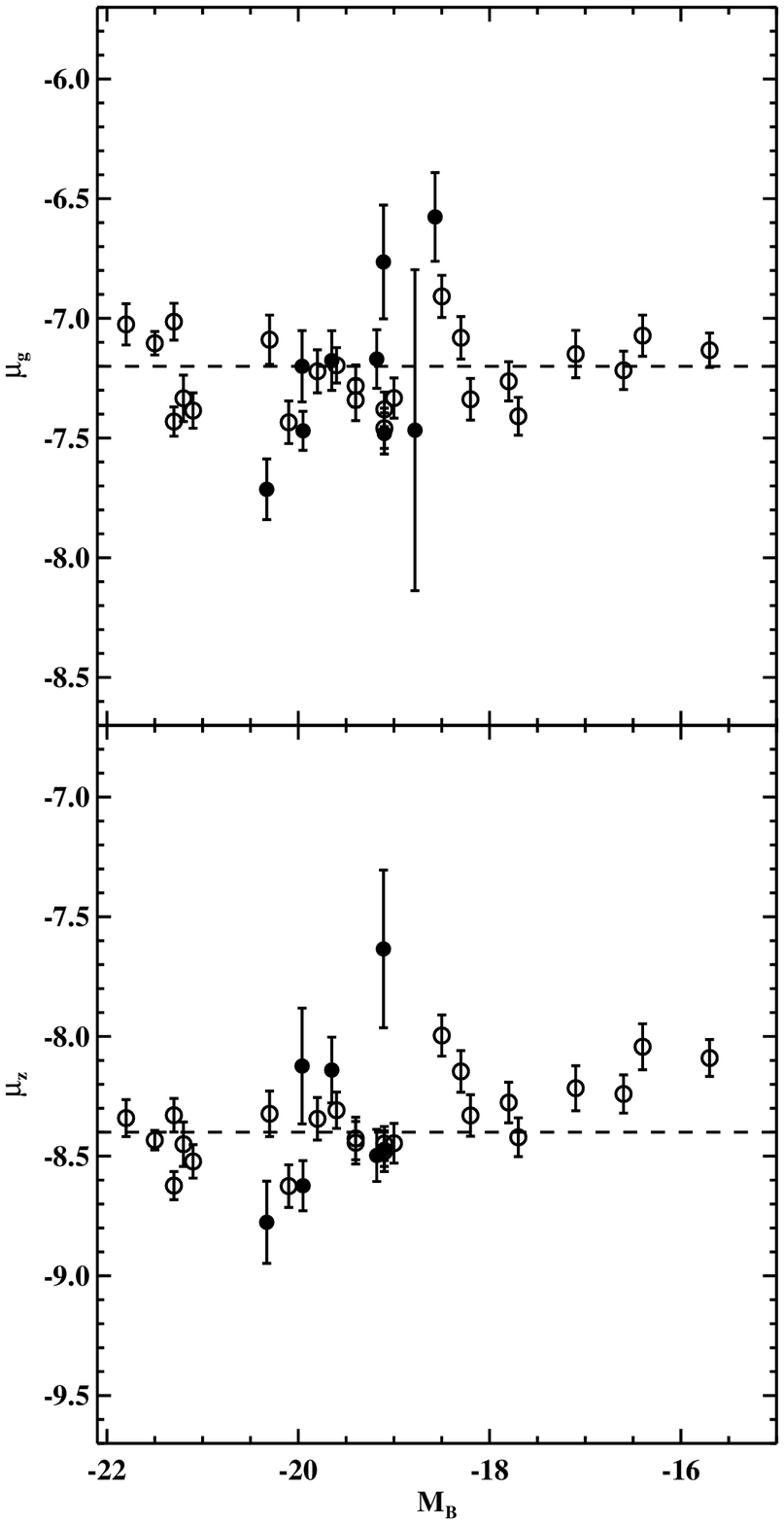} 
    \hspace{0.25in}
    \includegraphics[width=84mm]{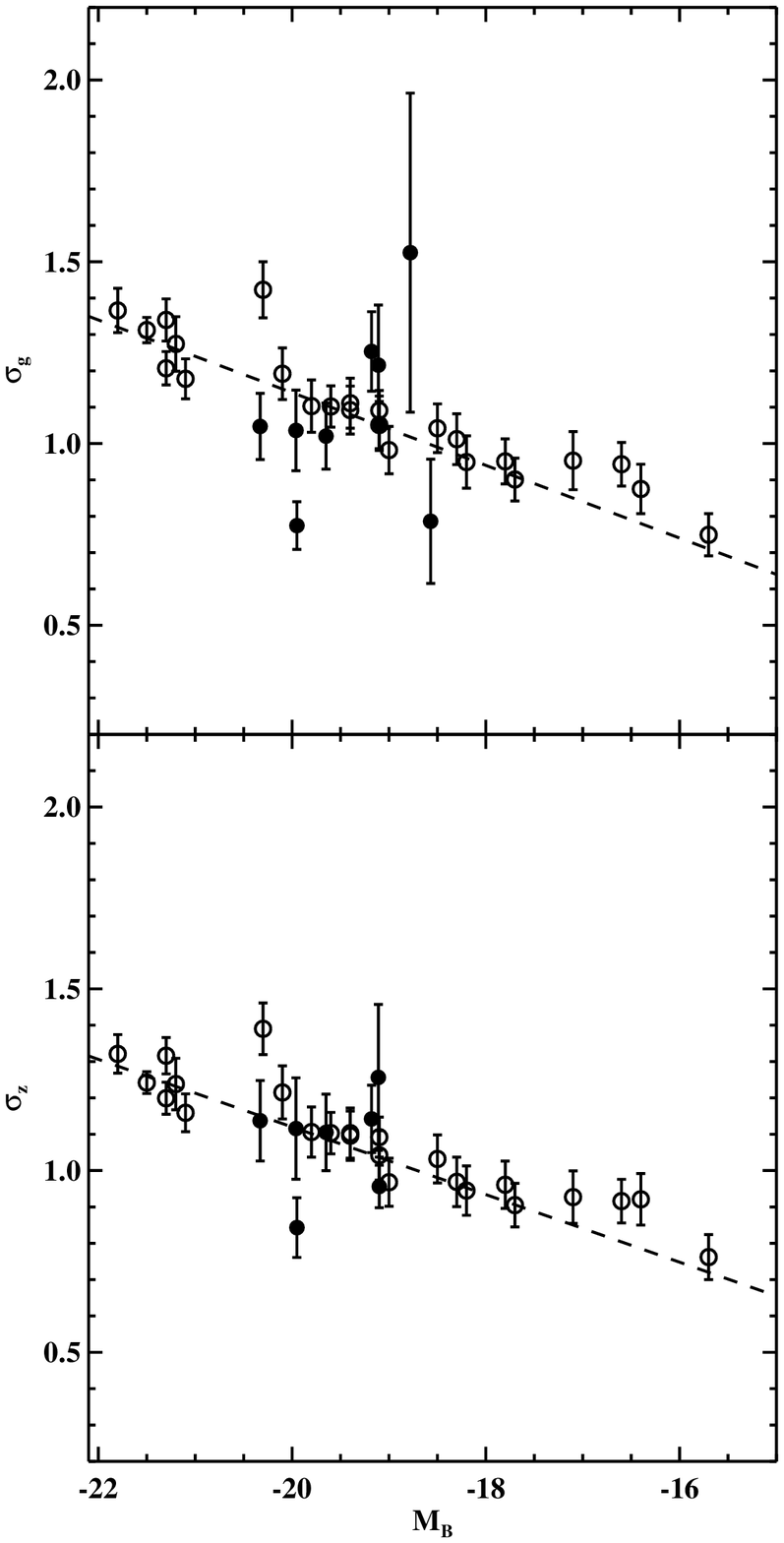} 
    } 
  }

  \vspace{5pt}

  \caption[Gaussian fit parameters of luminosity functions against host galaxy luminosity]{Gaussian fit parameters of luminosity functions against host galaxy luminosity. Turn-over magnitudes ({\it left}) and dispersions ({\it right}) in the $g$ and $z$ bands against host galaxy luminosity. The open circles represent the Gaussian fits to galaxy-binned GCLFs from the ACSVCS by \citet{jor07b}, whereas the results from this work are plotted with filled circles.  }
  \label{fig:gclf_com}

\end{figure*}

One might argue that because the majority of our sample galaxies are more distant than the Virgo cluster galaxies, and red GCs are more concentrated on the galaxy center, our detected GC samples would be less biased toward red GCs. Consequently, our red GC fraction could be systematically lower than that in the ACSVCS. Indeed NGC 3377, the nearest galaxy in our sample and closer than the Virgo cluster, has the highest red GC fraction. To test this hypothesis, we resampled GCs in NGC 3818, which is as twice distant as Virgo, within a circle with 50\arcsec radius that covers half of the ACS field of view. Therefore, the actual physical overage of this circle is similar to that of the ACSVCS. The new estimation of the red GC fraction within this circle is slightly higher by $\sim$0.02 mag than our initial estimate. In the case of NGC 7173, the original image contains 3 galaxies (NGC 7173, NGC 7174 and NGC 7176) and the actual coverage of NGC 7173 used for GC detection is only half of the full image. 
Considering NGC 7173 is as twice distant as Virgo, both physical coverages are similar, but the red fraction of GCs in NGC 7173 is still lower than that of the ACSVCS at the same galaxy luminosity. We therefore conclude that, although the ACSVCS samples are possibly biased to red GCs due to smaller physical coverage, the derived red GC fraction does not shift significantly and we still find a small systematic difference with the ACSVCS which may be related to the effects of local environment.

The shapes of the colour distributions in the ACSVCS are more varied than those in our sample. For instance, PJC06 found one single broad peak for the colour distributions of VCC 1664 and VCC 1619. They also found examples in which the red population completely dominates; red GC fractions of 0.84 in VCC1146 and NGC 4458 were measured. These types of ``abnormal'' colour distributions are not found in our smaller field sample.

To summarize the comparison of the properties of the GC systems in our sample with those of the ACSVCS, plots of the GC properties against host galaxy absolute magnitude (identical to Figures \ref{fig:mb_colmean_com}, \ref{fig:mb_colpeak_com}, and \ref{fig:mb_rf_com}) are shown in the first column of Figure \ref{fig:gal_gc_com}. The second and third columns in this figure are comparisons of the GC properties against the local galaxy number density (environment) and morphological type, respectively. In these two columns, only results from the ACSVCS within the range of the galaxy luminosity of our sample ($-18.4\geq M_{B}^{T}\geq -20.4$) are plotted so as to minimize any galaxy luminosity dependence. In the third column, the GC systems with the same host galaxy morphological type are randomly positioned within the morphological type bins. As can be seen in the second column, our sample is clearly distinct from the ACSVCS in terms of the galaxy environment. Because the GC system properties of both the ACSVCS and our sample are widely spread however, distinct differences with respect to the galaxy number density are not obvious in these plots. In the plots of the GC properties against morphological type (third column of Figure \ref{fig:gal_gc_com}) no obvious trends are observed with morphological type for the ACSVCS. However, for our sample, the GC systems in SB0/S0 type galaxies appear to have little/no bimodal colour distributions. In other words, the SB0/S0 galaxies have smaller red sub-populations when compared to early-type galaxies.

Our GCLFs have also been compared with those of the ACSVCS and the results are shown in Figure \ref{fig:gclf_com}. \citet{jor07b} fitted the GCLFs of 89 galaxies with both a Gaussian function and a Schechter function. They found that the GCLF dispersions were correlated with the galaxy luminosity, while the GCLF turn-over magnitudes were rather constant (in both bands). Later \citet{vil10} reanalysed \citet{jor07b}'s results for the Fornax cluster, finding  that the turn-over magnitude of GCLF gets fainter with decreasing galaxy luminosity. Using distance moduli from Table \ref{tab:gal_pro} and apparent turn-over magnitudes from Table \ref{tab:gclf}, we directly compared our results from \S \ref{sec:gclf_low} with the Gaussian fitting parameters of the galaxy-binned GCLFs reported by \citet{jor07b}. In Figure \ref{fig:gclf_com}, our results are overplotted on the fitting parameters from \citet{jor07b} and appear to be more widely spread than those from the ACSVCS. However, one must keep in mind that the ACSVCS results come from galaxy-binned GCLFs, which are less noisy than our individual GCLF fitting parameters. Using only our GCLF dispersions, it is hard to say whether or not there is a trend with galaxy luminosity because of the narrower range of galaxy luminosity in our sample. Our GCLF dispersions agree well with the ACSVCS, at least within the range of galaxy luminosities spanned by our sample.

\section{Discussion}
\subsection{Effect of environment on galaxy formation}

\begin{figure}
\begin{center}
\includegraphics[width=87mm]{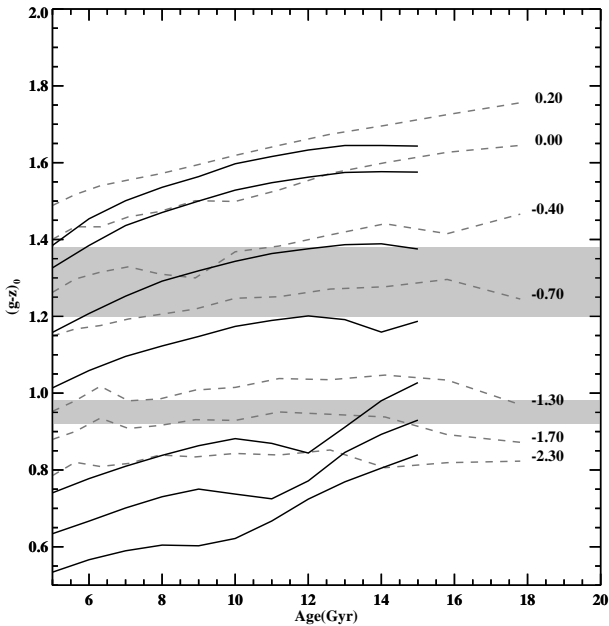}
\vspace{3mm}
       \caption[Evolution of g-z colour for various metallicities and ages]{Evolution of $g-z$ colour for various metallicities and ages. The grey dashed tracks are taken from simple stellar population models by \citet{gir06}, while the black ones are from YEPS. The metallicity noted at the end of each track is in units of [Fe/H]. The two stripes represent the colour ranges of red and blue peaks of GC systems in our sample galaxies. For old stellar populations ($>$ 8 Gyr), $(g-z)_{0}$ colour is mainly governed by metallicity rather than age.}
     \label{fig:ssp_gir}     
\end{center}
\end{figure}

From the analysis presented above, it appears that the mean colours of GC systems in field environments are slightly bluer than those in cluster environments at a given host galaxy luminosity. The simplest interpretation of this finding would be that these GC systems are either less metal-rich or younger than their counterparts in rich clusters (or {\it both} metal-poor {\it and} young). Simple stellar population models from Giraradi (2006, hereafter GIR06) and the Yonsei Evolutionary Population Synthesis (YEPS, http://web.yonsei.ac.kr/cosmic/data/YEPS.htm; C. Chung et al., in prep.) model were adopted in order to estimate the magnitude of these effects and compare them with our mean colours. The prediction of $(g-z)_{0}$ colour in various age and metallicity ranges based on the GIR06 and YEPS models is shown in Figure \ref{fig:ssp_gir}. The mean metallicity of the blue and red peaks shows good agreement with predictions based on the empirical transformation ($g-z$) to [Fe/H] (\citealt{pen06a}; \citealt{sin10}).

%It was first assumed that the blue and red sub-populations are coeval so that the colours purely depend on the metallicity. The metallicity range of each sub-population is denoted as a shaded area in Figure \ref{fig:ssp_gir}. Assuming an age of 11.2 Gyr, which is typical of metal-poor Galactic GCs (e.g. \citealt{sal02}; \citealt{ang05}), we estimate a metallicity range of $-1.89\leq [Fe/H] \leq  -1.59$ for the blue sub-populations and $-0.86\leq [Fe/H] \leq  -0.42$ for the red sub-populations. These ranges are in good agreement with the results of previous studies (Figure 2 in \citet{bro06}).

\begin{figure}
\begin{center}
\includegraphics[width=87mm]{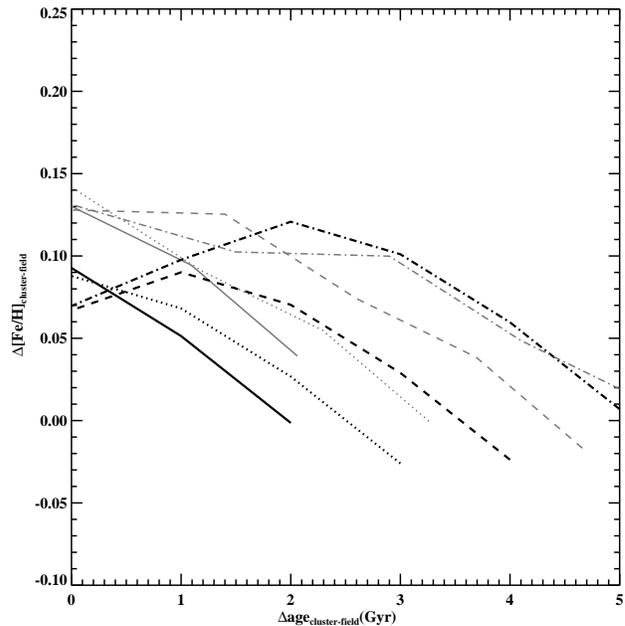}
\vspace{3mm}
       \caption[Metallcity differences of GCs between high density (cluster) and low density (field) environments]{Metallcity and age differences of GC systems between high density (cluster) and low density (field) environments. The black solid, dotted, dashed and dot-dashed lines represent 10, 11, 12 and 13Gyr of cluster GCs from the YEPS models respectively. The grey solid, dotted, dashed and dot-dashed lines are 10, 11.2, 12.6 and 14.1 Gyr models from GIR06 respectively. These model lines were derived from our results that the mean colours of GC systems in cluster galaxies are 0.05 mag redder than in field galaxies. }
     \label{fig:g_z_dfeh}     
\end{center}
\end{figure}

With only one colour ($g-z$), it is almost impossible to break the well-known age-metallicity degeneracy. Thus, an attempt was made to find any combination of metallicity and age differences that reproduce the mean colour offset between the high- and low-density GC systems. We first assumed an age range from 10 Gyr to 14 Gyr for the cluster environment GC systems and a minimum age of 8 Gyr for the field environment GC systems. We then took the mean colour of the cluster and field GC systems from Figure \ref{fig:mb_colmean_com} and derived the metallicity difference for a certain age combination from the YEPS and GIR06 models. The metallicity and age differences for cluster GC systems of various ages is shown in Figure \ref{fig:g_z_dfeh}. As an example, if the mean age of the cluster GC systems is 10 Gyr (the black solid line in Figure \ref{fig:g_z_dfeh}) and that of the field GC systems is 8 Gyr, then there seems to be no metallicity difference between the two GC systems according to the YEPS model. However, if the cluster and field GC systems are coeval, the cluster GC systems are, on average, more metal-rich by $\sim$ $0.10-0.15$ dex than the field GC systems. The YEPS model predicts a slightly lower metallicity difference than the GIR06 model when the two samples are coeval. This metallicity offset between the two samples disappears if the GC systems in the field galaxies are younger than those in the clusters. The age differences with no metallicity offset depend on the assumed ages of the GC systems in high-density environments and on the adopted simple stellar population model. However, the colour offset can be successfully explained by at least a $\sim$ 2 Gyr age offset when there is no metallicity difference. Therefore, we tentatively conclude that GC systems in low-density environments are either less metal-rich,  $\Delta[Fe/H]\sim0.10-0.15$, or  younger, $\Delta age>2$ Gyr, than those in high-density regions. Of course, any combination of age and metallicity within the above ranges can be a possible solution, as seen in Figure  \ref{fig:g_z_dfeh}. 

Metallicity difference between the low- and high-density GC systems has been also derived from a fully empirical colour-metallicity relation by \citet{bla10}, in which a quartic polynomial is fit to compiled data from PJC06. Based on the relation, the colour offset is found to correspond to $\Delta[Fe/H]\sim0.12$. Although the empirical relation might contains age effects, the metallicity difference is well consistent with that from the SSP models.

Even though we managed to detect the colour offset of GC systems in low- and high-density environments, and quantify the age and metallicity difference using the simple stellar population models and the empirical colour-metallicity relation, the detected colour offset ($\Delta(g-z)_{0}\sim0.05$) is not comparable with the colour range due to host galaxy luminosity. For example, within our sample, the largest difference in mean colour is $\Delta(g-z)_{0}\sim0.13$ (see Table \ref{tab:kmm}) and for the entire ACSVCS sample, the mean GC system colour has a range of $0.76\leq(g-z)_{0}\leq1.25$ (\citealt{pen06a}). Therefore, stellar populations in GC systems are mainly governed by their host galaxy mass (luminosity) and environmental effects are less important in determining the star formation history of early-type galaxies and their GC systems.

There have been a few previous attempts to detect differences in GC properties depending on the environments to which their host galaxies belong. \citet{geb99} used 50 mostly early-type galaxies from the HST/WFPC2 archive and found no correlation between the colour peaks and the host galaxy properties in 15 field sample galaxies. However, these researchers found a strong correlation of GC system peak colours versus galaxy velocity dispersion and Mg2 index in their cluster galaxy sample. In our field galaxy results, a strong relationship between the mean colour and the galaxy luminosity was certainly observed (see Figure \ref{fig:mb_col}). One possible reason why \citet{geb99} did not find such a relationship in their field galaxy sample is that their galaxies were too luminous ($M_{V}\la-20$) and only small differences are to be expected for luminous ($M_{B}\la-21$) early-type galaxies (e.g., \citealt{bau96}; \citealt{col00}). Another possible reason is that the photometric accuracy of the HST/WFPC2 data was not sufficient to detect the correlation. In fact, the present work and the research of \citet{geb99} have two field galaxies in common: NGC 1426 and NGC 3377. When compared to \citet{geb99}, we detected twice as many GCs in these two galaxies. \citet{lar01} also briefly investigated the dependency of colour bimodality on the environment in an independent sample of 17 early-type galaxies obtained from HST/WFPC2, but did not find any correlation. As can be seen from our results, the colour bimodality of GC systems mainly depends on the galaxy luminosity (mass), although this trend disappears at slightly higher galaxy luminosities in low-density regions.

\citet{pen08} found a large spread in the specific frequency ($S_{N}$) of dwarf galaxies from the ACSVCS. Almost all dwarfs with a high $S_{N}$ were located within a projected radius of 1 Mpc from M87 (Figure 4 in \citet{pen08}), whereas no $S_{N}$ difference between the central region and the outskirts of the Virgo cluster was found for intermediate luminosity galaxies (i.e. similar to those in our low-density sample). While \citet{pen08} found a dependency of $S_{N}$ for the dwarfs on the environment, the cumulative GC colour distributions of these two groups show no obvious difference. It is possible that the effect of environment on the colour distribution is so subtle that it could not be detected within a cluster environment such as Virgo.

Regarding the general dependence of galaxy formation on environment, many observational and theoretical studies have been conducted without including constraints from GC system data. It is now well-established that late-type galaxies are biased to low-density environments and giant elliptical galaxies are preferentially located in galaxy clusters (\citealt{dre80}). Cluster galaxies also exhibit lower star formation rates than field galaxies at a given redshift, luminosity, and bulge-to-disk mass ratio (\citealt{bal98}; \citealt{lew02}; \citealt{go03}). Early-type galaxies in low-density environments also appear to be younger ($\Delta age\sim$ 2 Gyr) and more metal-rich ($\Delta [Fe/H]\sim$ 0.1 dex) when compared to their counterparts in dense environments (\citealt{kun02}; \citealt{tho05}), although \citet{cle06} found no environmental influence on the metallicity (but still found younger ages for field galaxies). These observational findings are contradictory to the semi-analytic model of hierarchical galaxy formation (e.g. \citealt{bau96}, \citealt{col00}), which predicts a larger spread of age and metallicity and younger ages and lower metallicities (on average) in low-density environments, especially in low-luminosity early-type galaxies (\citealt{bau96}, \citealt{col00}).

Our results for the GC systems are somewhat inconsistent with those of \citet{kun02} and \citet{tho05}, who found higher metallicities for galaxies in field environments (which is consistent with the semi-analytic models from \citet{bau96} and \citet{col00}). The question remains as to whether GC formation history is different from field star formation history or does the environment have different effects on GCs and field star formation. \citet{har02} have suggested that GCs formed somewhat earlier than most field stars based on their discovery of a higher $S_{N}$ at a low metallicity in NGC 5128. This argument is also supported by \citet{pen08}, who estimated GC formation rates from the Millennium Simulation (\citealt{spr05}) and showed that cluster formation peaks earlier than field star formation. Such results would explain why GCs are more metal-poor than field stars in early-type dwarf galaxies. The higher $S_{N}$ of dwarf ellipticals in the Virgo cluster center was successfully explained by the higher peak star formation rate (SFR) and star formation surface density ($\Sigma_{SFR}$) in the cluster central region when compared to the outskirts and a model where GC formation rate $CFR\propto SFR(\Sigma_{SFR})^{0.8}$.

As described in \S \ref{sec:com_vcs}, the colour of GC systems is mainly determined by the mass of the host galaxy in both cluster and field environments. Subtle environmental differences also exist; the mean metallicities of GC systems in field galaxies are slightly lower than those in cluster galaxies at a given host galaxy luminosity (mass), while the fraction of red clusters is higher in dense environments. From our findings, it is suggested that GC systems in field galaxies form later than those in cluster galaxies and/or have a lower metallicity. Furthermore, we expect that in dense environments, cluster formation history is complicated by disturbing or interacting neighbor galaxies. This is reflected by the larger degree of variation in the shapes of the colour distributions that are found in dense environments. While \citet{pen08} found no difference in the $S_{N}$ of low-luminosity ellipticals (intermediate luminosity according to their terminology) in their Virgo cluster sample, in the future it would be worth comparing the $S_{N}$ of our sample with that of the ACSVCS.

\subsection{Metal-poor GC colour--galaxy luminosity relation}
 \label{sec:mpgc}
 
 The existence (or non-existence) of a correlation between the peak of the blue GC colours and the host galaxy luminosity can have implications for different galaxy formation scenarios. Major merger \citep{ash92} and accretion models \citep{cot98} would have difficulty explaining this correlation because, in the merger model, metal-poor GCs in two lower equal mass spiral galaxies are still a primary resource of metal-poor GCs in the final ellipticals. In the accretion model, metal-poor GCs come from dwarf galaxies, so the mean metallicities of GCs in dwarf galaxies and metal-poor GCs in ellipticals are more or less the same. Many early studies found a strong correlation between the colours of metal-rich GCs and the host galaxy luminosity, but failed to detect any evidence of a similar correlation with the mean colours of metal-poor GCs (e.g., \citealt{for97}; \citealt{kun01a}; \citealt{for01}). However, \citet{lar01} found a weak correlation between the mean colours of metal-poor GCs and the galaxy absolute magnitude in their HST/WFPC2 sample.  \citet{bur01} also suggested that such a relationship may exist. \citet{lot04} found that the slope of the GC peak of their dwarf elliptical sample ($M_{B}\geq -18$), when plotted against galaxy luminosity, is consistent with that of the GC blue peak in early-type galaxies found by \citet{lar01}.
 
 Recent studies of the correlation between metal-poor GCs and host galaxy luminosity have been carried out by \citet{str06} and \citet{pen06a}. The former compiled blue GC peak data for early-type galaxies and local spirals from various sources (\citealt{lar01}; \citealt{kun01a}, \citeyear{kun01b}; \citealt{har96}; \citealt{bar00}; \citealt{ols04}) and found a significant ($>5 \sigma$) correlation between the blue colour peak and galaxy luminosity. \citet{pen06a} also detected this blue peak trend in homogenous ACSVCS data, but it was not as strong as that for the red peak. Using their own relationship between $g-z$ colour and [Fe/H], \citet{pen06a} found that the slope of the blue peak metallicity-galaxy luminosity relationship is even steeper than that found in the colour-galaxy luminosity plane due to the steeper colour-metallicity relationship in the metal-poor region (see Figure 12 in  \citet{pen06a}). However, they noted that this blue slope varies depending on the adopted colour-metallicity relationship (simple stellar population model or empirical transformation) by a factor of $\sim3$. 
 %Meanwhile, \citet{kun08} argued that correlation between mean size of metal-poor GCs and galaxy mass can possibly result in blue peak of GCs and host galaxy luminosity relation in high-resolution HST images.

In Figure 2 of \citet{bro06}, the peak metallicities of GC sub-populations are plotted against galaxy luminosity with the peak positions taken from \citet{str04} and \citet{str06}. In this plot, the peak metallicities of both metal-poor and metal-rich GCs appear to be correlated with host galaxy luminosity with similar slopes. The metal-poor GC trend can be accounted for with an in-situ scenario \citep{for97} in which metal-poor GCs formed in the Universe first, and after a sudden truncation (possibly by reionization) metal-rich GCs then formed along with the bulk of the field stars in galaxies. 

From our results on the blue peak colours alone, it is not clear whether or not the blue peak colours are correlated with the host galaxy luminosity because of the narrow galaxy luminosity range in our sample (only $\sim2$ mag range in $M_{B}$ compared to a range of $\sim6$ mag in the ACSVCS). To observe any trend in the blue peak colour with galaxy luminosity, we would need to study GC systems in fainter galaxies ($M_{B}\geq -18$, i.e., dwarf ellipticals) in low-density environments (similar to those studied by \citet{lot04}). This would serve to strengthen the correlation between the GC blue peak and galaxy luminosity in the higher density Virgo and Fornax clusters. It is thus unclear whether the slope of the blue (metal-poor) GC peak in field environments is different from that in clusters. In fact, \citet{str04} predict that the slope of metal-poor GC metallicity-galaxy luminosity plots in dense environments should be steeper than that in low-density environments if cosmic reionization caused a quenching of blue GC formation. The role of cosmic reionization in GC formation is, however, uncertain. Whether cosmic reionization can really truncate GC formation and how efficient GC formation is prior to the reionization is still controversial (e.g. \citealt{mur10}; \citealt{gri10}).

\section{Conclusions}

\begin{itemize}

\item High spatial resolution images of 10 early-type galaxies in low-density (field) environments have been obtained using the HST/ACS in the F457W and F850LP bands. The properties of the GC systems associated with these galaxies and their connection with host galaxy properties have been investigated. Our results have been compared with those of the ACS Virgo Cluster Survey (ACSVCS) in order to study the role of the environment in galaxy formation.

\item The GC system properties of our low-density sample exhibit trends with respect to the host galaxy luminosity that are similar to those in clustered environments. There are more total GCs and more red GCs in luminous galaxies, while the mean colour and the colour of the red peak GCs are strongly correlated with host galaxy luminosity. Colour bimodality becomes less clear with a decrease in galaxy luminosity.

\item The mean colours in low-density (field) regions appear to be slightly bluer than those around galaxies of equivalent luminosities in high-density regions. The fraction of red GCs in field galaxies is found to be lower than that in clustered galaxies of similar luminosities. When compared to the trend observed in the ACSVCS, the slope of the red peaks against host galaxy luminosity is steeper. In other words, colour bimodality disappears more quickly as galaxies get fainter, in low density environments.

\item Luminosity functions of the GCs in most of our sample galaxies are well-fit by a Gaussian function whose fitting parameters (turn-over magnitude and dispersion) agree well with those of the ACSVCS. We find no independent evidence for a trend of dispersion in the fitted Gaussian luminosity functions against galaxy luminosity.

\item We have investigated the possible origins of the unexpected excess of GCs in NGC 3377 near $g$ $\sim25.5$, which deviate from a normal Gaussian luminosity function. The origin of this population is still not fully understood, but some of clusters are likely to be background galaxies from visual inspection and some are possibly diffuse star clusters.

\item At a given luminosity, the mean colours of our GC systems in low-density environments are slightly bluer ($\Delta(g-z)_{0}\sim0.05$) than their counterparts in high-density environments. By assuming various combinations of age and employing simple stellar population models, this colour offset corresponds to a metallicity difference of $\Delta[Fe/H]\sim0.10-0.15$ or an age difference of at least $\Delta age\sim2$ Gyr on average. Therefore, GCs in field galaxies appear to be either less metal-rich or younger than those in cluster galaxies of the same luminosity.

\item Whilst a correlation between the blue peak colour of GCs and host galaxy luminosity was found in the Virgo cluster (ACSVCS), no such correlation was detected in our low-density galaxy sample. It is not clear whether the absence of such a trend is because our galaxy luminosity range is simply too narrow or because intrinsically the slope of the relationship depends on the environment. For a more definitive determination, observations of GCs in dwarf ellipticals in low-density environments are needed.

\item The greater variation in the shapes of the colour distributions for GC systems in the Virgo cluster sample could imply that more complex galaxy formation processes (e.g., interactions/harassment with adjacent galaxies) are taking place in the galaxy clusters. The higher fraction of red GCs in cluster galaxies also supports this possibility.

\item Although we found that the galaxy environment has a subtle effect on the formation and metal enrichment of GC systems, host galaxy mass is the primary factor that determines the stellar populations of both the GCs and the galaxy itself. The processes which determine GC formation must therefore be largely common across a range of galaxy environments. This remains a challenge for GC formation theories.

\end{itemize}

\vspace{5mm}
\section*{Acknowledgments}

JC acknowledges support by the Yonsei University Research Fund of 2009.
SJY acknowledges support from Basic Science Research Program (No. 2009-0086824)
through the National Research Foundation (NRF) of Korea grant
funded by the Ministry of Education, Science and Technology (MEST),
and support by the NRF of Korea to the Center for Galaxy Evolution Research
and the Korea Astronomy and Space Science Institute Research Fund 2011. SEZ acknowledges support for this work from HST
grant HST-GO-10554.01. A. Kundu acknowledges support from HST archival program HST-AR-11264.

\end{document}